\documentclass[12pt,a4paper,final,longbibliography]{iopart}
\usepackage{iopams}
\usepackage{graphicx}
\usepackage[breaklinks=true,colorlinks=true,linkcolor=blue,urlcolor=blue,citecolor=blue]{hyperref}

\expandafter\let\csname equation*\endcsname\relax
\expandafter\let\csname endequation*\endcsname\relax
\usepackage{bm, amssymb, color}

\usepackage{amsmath}
\usepackage[caption=false]{subfig}
\usepackage{graphicx}
\usepackage{amsfonts}
\usepackage{mathtools}
\usepackage{float}
\usepackage{epsfig}
\usepackage{color}
\usepackage{gensymb}
\usepackage{multirow}
\usepackage[mathscr]{euscript}
\usepackage{pifont}
\usepackage{amsmath,soul}
\usepackage[export]{adjustbox}
\usepackage{array}
\usepackage{cases}
\usepackage{makecell}
\usepackage{physics}
\usepackage{yhmath}
\usepackage{enumitem}
\setstcolor{blue}

\setul{0}{0.3ex}

\newcommand{\spD}[1]{\fn{\tilde{\chi}_{_V}}{#1}}

\newcommand{\R}{\mathbb{R}}
\newcommand{\vv}[1]{\fn{\sigma_{_V} ^2}{#1}}
\newcommand{\fn}[2]{\mathinner{#1\mathopen{\left(#2\right)}}}
\newcommand{\vect}[1]{{\bf #1}}

\begin{document}

\title{Local Order Metrics for Two-Phase Media Across Length Scales}

\author{Salvatore Torquato$^{1}$, Murray Skolnick$^{2}$ and Jaeuk Kim$^{3}$}%

\address{$^1$ Department of Chemistry, Department of Physics, Princeton Institute for the Science and Technology of Materials, Program in Applied and Computational Mathematics, Princeton University, Princeton, New Jersey 08544, USA}

\address{$^2$ Department of Chemistry, Princeton University, Princeton, New Jersey 08544, USA}

\address{$^3$ McKetta Department of Chemical Engineering, University of Texas at Austin, Austin, Texas 78712, USA}

\date{\today}

\begin{abstract}

The capacity to devise order metrics for microstructures of multiphase heterogeneous media is a highly challenging task,
given the richness of the possible geometries and topologies of the phases that can arise. This investigation
 initiates a program to formulate order metrics to characterize the
degree of order/disorder of the microstructures of two-phase media in $d$-dimensional Euclidean space $\mathbb{R}^d$ across length scales.
In particular, we propose the use of the local volume-fraction variance $\sigma^2_{_V}(R)$ associated 
with a spherical window of radius $R$ as an order metric. We determine $\sigma^2_{_V}(R)$ as a function of $R$ for 22 different models
across the first three space dimensions, including both hyperuniform and nonhyperuniform systems with varying
degrees of short- and long-range order. We find that the local volume-fraction variance as well as asymptotic coefficients and integral measures derived from it provide reasonably robust and sensitive order metrics to categorize disordered and ordered two-phase media across all length scales.

\end{abstract}

\date{\today}
\maketitle
\section{Introduction}

Heterogeneous multiphase media and materials abound in nature and synthetic situations. Examples of such materials
include composites, porous media, foams, cellular solids, colloidal suspensions, polymer
blends, geological media, and biological media \cite{Ch79,To02a,Mi02,Sa03,Bu07}.
While the study of order metrics to characterize the degree of order of point configurations has been
a fruitful endeavor \cite{To18b},
it is much more challenging to devise such order metrics  to describe the microstructures of multiphase media for two reasons. First, the 
geometries and topologies of the phases are generally much richer and more
complex than point-configuration arrangements. Second, one
must determine characteristic microscopic length scales that are broadly applicable for
the multitude of possible two-phase media microstructures.

This paper initiates a program to formulate order metrics to characterize the
degree of order/disorder of the microstructures of two-phase media in $d$-dimensional Euclidean space $\mathbb{R}^d$ across length scales.
In particular, we propose the use of various measures of volume-fraction fluctuations
within a $d$-dimensional spherical window of radius $R$ as order metrics.
Such fluctuations are known to be of importance in a variety
of problems, including  the study of noise and granularity of photographic
images \cite{Bay64,Lu90a}, transport through composites and porous
media \cite{To20},  the properties of organic coatings \cite{Fi92}, the fracture of composite materials \cite{Bo97},
and the scattering of waves in heterogeneous media \cite{Ke64b,Ki20a,Da21}.

For concreteness, we focus on two-phase media in $\mathbb{R}^d$ in this work, but we note that the generalization of our results to $n$-phase
media is straightforward. The global volume fractions of phases 1 and 2 are denoted by $\phi_1$ and $\phi_2$, respectively,
where $\phi_1+\phi_2=1$. At a local level, the phase volume fraction fluctuates. 
The simplest measure of volume-fraction fluctuations is the {\it local volume-fraction variance}
$\sigma^2_{_V}(R)$ (see Fig. \ref{cartoon}), which
can be expressed in terms of the autocovariance function $\chi_{_V}({\bf r})$  \cite{Lu90b,Za09} (defined in Sec. \ref{back}):

\begin{figure*}[!t]
    \subfloat[\label{sampling-debye}]{\includegraphics[width=0.33\textwidth]{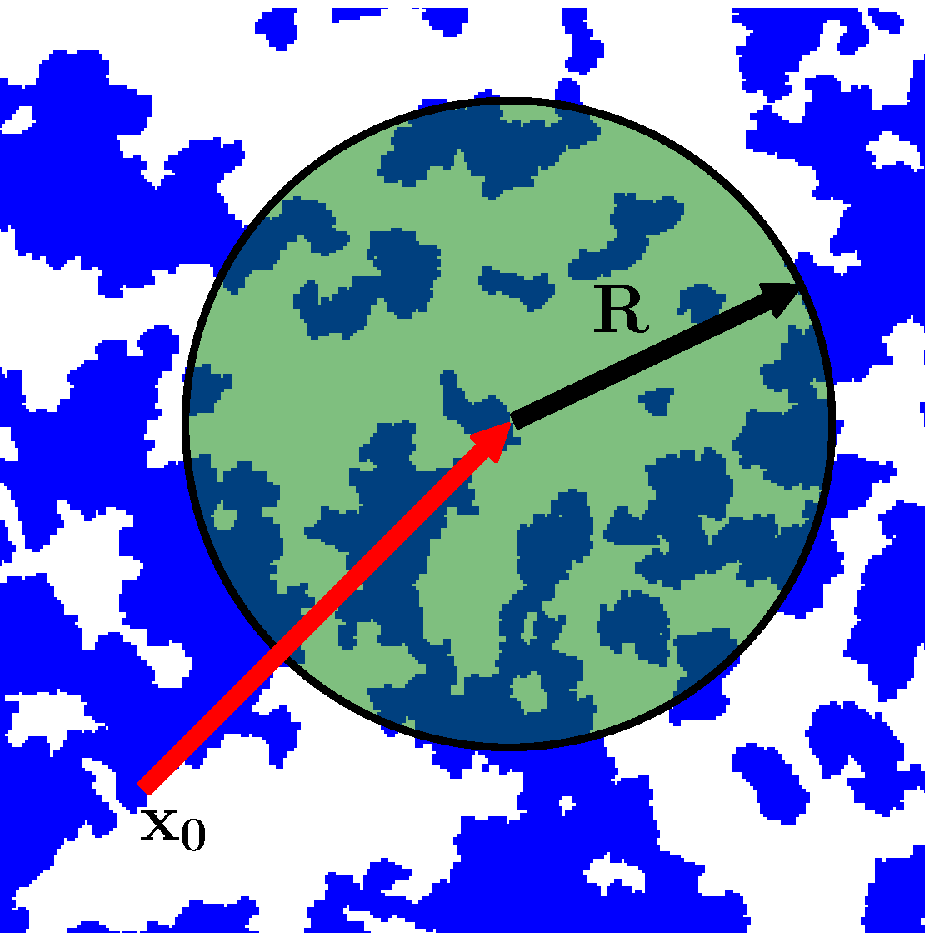}}
    \subfloat[\label{sampling-stealthy}]{\includegraphics[width=0.33\textwidth]{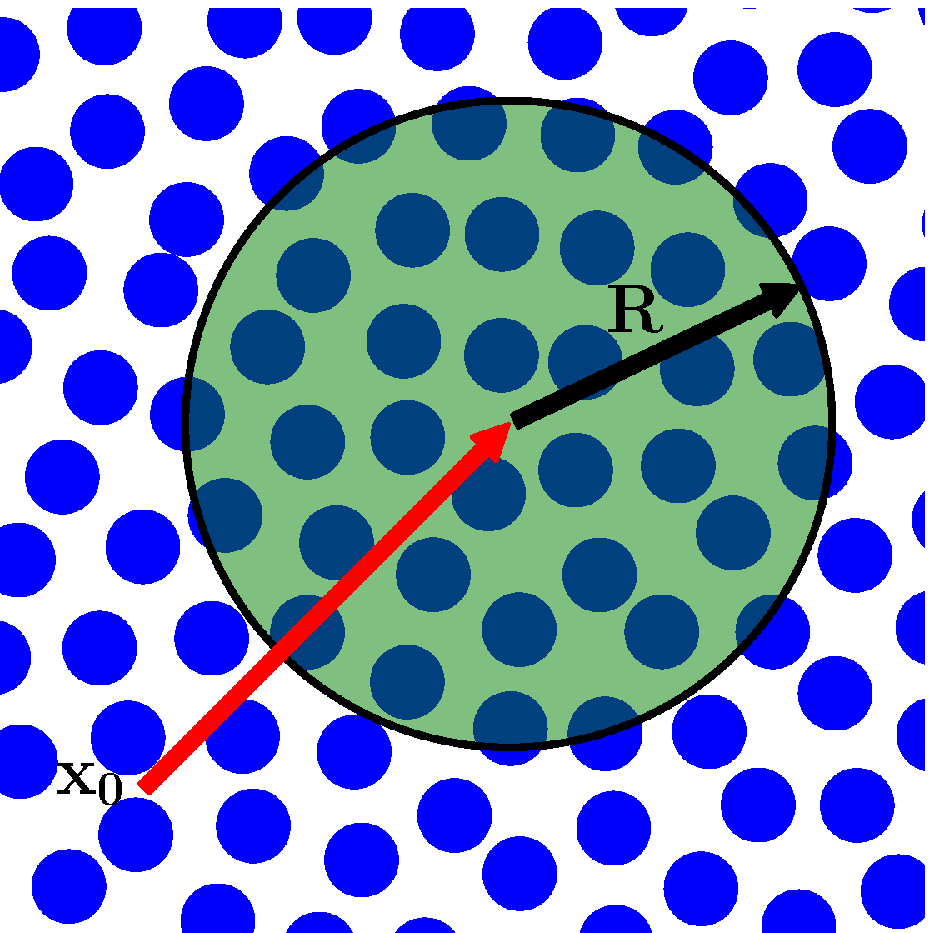}}
    \subfloat[\label{sampling-ordered}]{\includegraphics[width=0.33\textwidth]{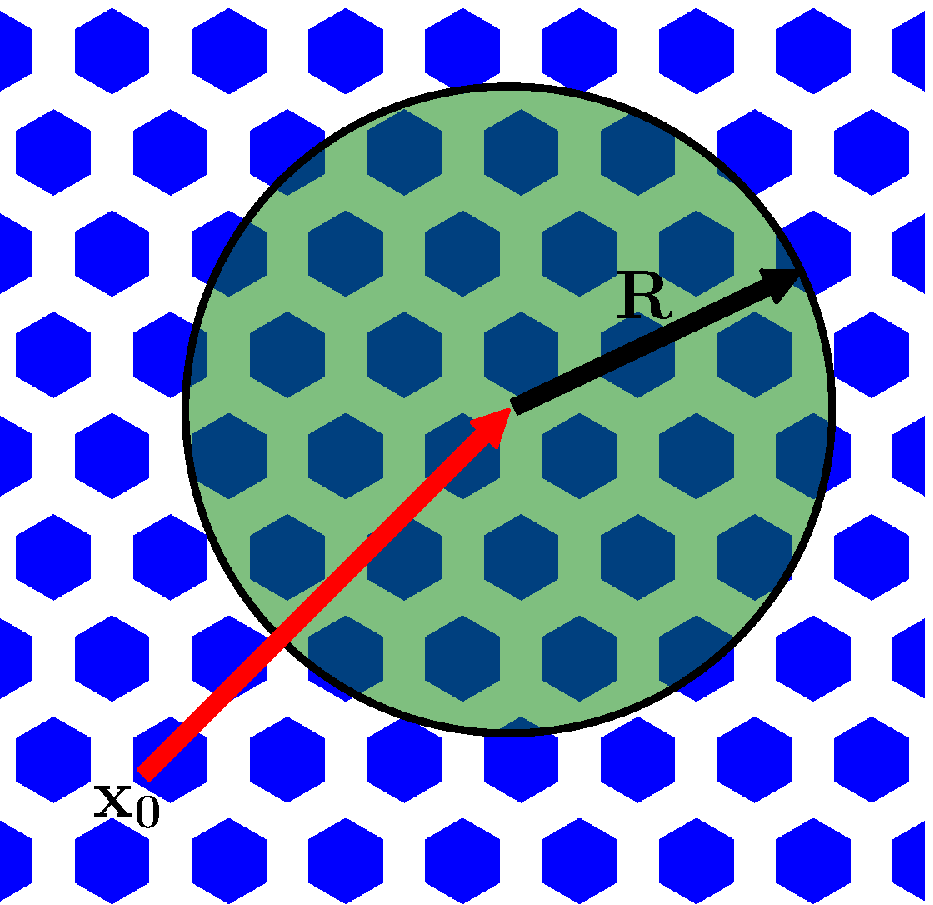}}
	\caption{ Schematics indicating a circular observation window of radius $R$ in two dimensions, and its centroid $\bf x_0$ for a general two-phase disordered nonhyperuniform heterogeneous medium (a), a disordered hyperuniform two-phase heterogeneous medium (b), and an ordered two-phase heterogeneous medium (c), as adapted from Ref. \cite{To18a}. In each of these examples, the phase volume fraction  within a window fluctuates as the window centroid varies. While the local variance $\sigma_{_V}^2(R)$ for the nonhyperuniform medium decays like $1/R^2$, it decays like $1/R^3$ in both the disordered hyperuniform and periodic examples. }\label{cartoon}
\end{figure*}

\begin{equation}
\sigma_{_V}^2(R) = \frac{1}{v_1(R)} \int_{\mathbb{R}^d} \chi_{_V}(\mathbf{r})\, \alpha_2(r; R) d\mathbf{r},
\label{sig-direct}
\end{equation}
where 
\begin{equation}
v_1(R)=\frac{\pi^{d/2} R^d}{\Gamma(d/2+1)}
\label{v1}
\end{equation} 
is the volume of a $d$-dimensional sphere of radius $R$, $\Gamma(x)$ is the gamma function
and  $\alpha_2(r;R)$ is the intersection volume of two spherical windows of radius $R$ separated by a distance
$r$ divided by the volume of a window. The quantity $\alpha_2(r;R)$ is known analytically
in any space dimension \cite{To06b}. Note that $\sigma^2_{V}(R=0)=\phi_1\phi_2$ \cite{Lu90b}, which can be proved to be an upper bound
on the local variance, i.e., $\sigma^2_{V}(R) \le \phi_1\phi_2$ for all $R$. In addition to the direct-space representation (\ref{sig-direct}), the local variance $\vv{R}$  has the following Fourier-space representation in terms of the spectral density $ \spD{\vect{k}}$ \cite{To18a, Za09}: 
\begin{align}
\sigma_{_V}^2(R) & = \frac{1}{\fn{v_1}{R}(2\pi)^d} \int_{\R^d} \spD{\vect{k}} \fn{\tilde{\alpha}_2}{k;R} \dd{\vect{k}}
, \label{sig-Fourier}
\end{align}
where $\bf k$ is the wavevector, $k\equiv |{\bf k}|$ is the wavenumber,
\begin{equation} \label{alpha}
\fn{\tilde{\alpha}_2}{k;R} \equiv 2^d \pi^{d/2} \fn{\Gamma}{d/2+1}\frac{\qty[\fn{J_{d/2}}{kR}]^2}{k^d}
\end{equation} 
is the Fourier transform of $\alpha_2(r;R)$ \cite{To03a}, and $J_{\nu}(x)$ is the Bessel function of the first kind of order $\nu$ (see Sec. \ref{specs} for plots of $\fn{\tilde{\alpha}_2}{k;R}$ for $d=1,2,$ and $3$).
The spectral density $\spD{\vect{k}}$ is the Fourier transform of the autocovariance function $\chi_{_V}(\bf r)$
and is directly related to the scattering intensity \cite{De57}.

The large-$R$ behavior of $\sigma^2_{_V}(R)$ is at the heart of the hyperuniformity concept.
Hyperuniform two-phase media are characterized by
an anomalous suppression of volume-fraction fluctuations relative to garden-variety disordered media \cite{Za09,To18a}
and can be endowed with novel  properties  \cite{Fl09b,De16,Fr17,To18a,Bi19,Go19,Le19a,Ki20a,Sh20,Zh20b,Ch21,Ch21b,Ni21,To21a,Zh21,Yu21,Du22}.
Specifically, a hyperuniform two-phase system is one in which $\vv{R}$ decays faster than $R^{-d}$ in the large-$R$ regime \cite{Za09, To18a}, i.e.,  
\begin{equation}\label{eq:HU-condition2}
\lim_{R\to\infty}R^d \vv{R} = 0.
\end{equation}
Equivalently, a hyperuniform medium is one in which the {\it spectral density} $\spD{\vect{k}}$ goes to zero
as $|{\bf k}|$ tends to zero \cite{Za09, To18a}, i.e., 
\begin{equation}\label{eq:HU_condition}
\lim_{\abs{\vect{k}}\to 0 }\spD{\vect{k}} = 0.
\end{equation}

The hyperuniformity concept has led to a unified means to classify equilibrium and nonequilibrium states
of matter, whether hyperuniform or not, according to their large-scale fluctuation characteristics.
Suppose the spectral density has the following power-law behavior as $|\bf k|$ tends to zero:
\begin{equation}
{\tilde \chi}_{_V}({\bf k})\sim |{\bf k}|^\alpha \quad  (|\mathbf{k}|\to 0),
\label{spec-scaling}
\end{equation}
where $\alpha$ is an exponent that specifies whether the medium is hyperuniform or not. 
In the case of hyperuniform two-phase media,  $\alpha>0$, and it has been shown \cite{Za09,To18a} that  there are three different scaling regimes (classes) that describe the associated large-$R$ behaviors of the volume-fraction variance
\begin{align}  
\sigma^2_{_V}(R) \sim 
\begin{cases}
R^{-(d+1)}, \quad\quad\quad \alpha >1 \qquad &\text{(Class I)}\\
R^{-(d+1)} \ln R, \quad \alpha = 1 \qquad &\text{(Class II)}\\
R^{-(d+\alpha)}, \quad 0 < \alpha < 1\qquad  &\text{(Class III).}
\end{cases}
\label{sigma-hyper}
\end{align}
Classes I and III are the strongest and weakest forms of hyperuniformity, respectively.
Class I media include all crystal structures, many quasicrystal structures and exotic disordered media \cite{Za09,To16b,To18a}.
 
By contrast, for any nonhyperuniform two-phase system, it is straightforward to show,
using a similar analysis as for point configurations \cite{To21c}, that
the local variance has the following large-$R$ scaling behaviors:
\begin{align}  
\sigma^2_{_V}(R) \sim 
\begin{cases}
R^{-d}, & \alpha =0 \quad \text{(typical nonhyperuniform)}\\
R^{-(d+\alpha)}, & -d <\alpha < 0 \quad \text{(antihyperuniform)}.\\
\end{cases}
\label{sigma-nonhyper}
\end{align}
For a  ``typical" nonhyperuniform system, ${\tilde \chi}_{_V}(0)$ is bounded  \cite{To18a}. For {\it antihyperuniform} media,
${\tilde \chi}_{_V}(0)$ is unbounded, i.e.,
\begin{equation}
\lim_{|{\bf k}| \to 0} {\tilde \chi}_{_V}({\bf k})=+\infty,
\label{antihyper}
\end{equation}
and hence are diametrically opposite to hyperuniform systems.
Antihyperuniform media include systems at thermal critical points (e.g., liquid-vapor and magnetic critical points) \cite{St87b,Bi92}, fractals \cite{Ma82}, disordered non-fractals \cite{To21b},
and certain substitution tilings \cite{Og19}.

In this paper, we propose the use of the local volume-fraction variance $\sigma^2_{_V}(R)$ as an order metric for disordered and ordered two-phase media across all length scales by tracking it as a function 
of $R$. Specifically, for any particular value of $R$, the lower the volume-fraction fluctuations as measured by $\vv{R}$, the greater the degree of order.
We study this order metric and integral measures derived from it across length scales for a large family of models across
the first three space dimensions, including both hyperuniform and nonhyperuniform systems with varying
degrees of short- and long-range order. This constitutes 22 different two-phase models across
the first three space dimensions. Examination of the same model across dimensions enables
us to study the effect of dimensionality on the ranking of order across dimensions. For almost all one-dimensional (1D) models and some two-dimensional (2D) and three-dimensional (3D) models, we obtain exact closed-form formulas for their pair statistics and local variances. We find that the local volume-fraction fluctuations,
as measured by the magnitude of $\sigma^2_{_V}(R)$ for a particular value of the window radius $R$, provide a reasonably robust way to rank order different two-phase media at a common global volume fraction. We also calculate the implied coefficients 
multiplying the large-$R$ scaling of the variance for class I hyperuniform media [cf. (\ref{sigma-hyper})]
and that of typical nonhyperuniform media [cf. (\ref{sigma-nonhyper})]. 
The calculation of such large-$R$ asymptotic coefficients was only recently carried
out, but primarily for certain 2D ordered structures \cite{Ki21}.

In Sec. \ref{back}, we present necessary definitions and background material. In Sec. \ref{Fourier}, we derive a useful Fourier-space
 representation of a large-$R$ asymptotic coefficient that we employ in subsequent sections.
Brief descriptions of the two-phase models across the first three space dimensions and their 
corresponding relevant structural characteristics are given in Sec. \ref{models}.
In Sec. \ref{results-1}, we present the local variance as well as integral measures derived from it as order metrics. In Sec. \ref{results-2}, we present the major results for the local variance for all of the models. Finally, we make concluding remarks in Sec. \ref{conclusions}.

\section{Definitions and Background}
\label{back}

\subsection{Correlation Functions}

A two-phase medium is fully statistically characterized by the $n$-point correlation functions \cite{To02a},
defined by
\begin{equation}
S_{n}^{(i)} \left( \mathbf{x}_1, ..., \mathbf{x}_n \right) \equiv
\left \langle {\cal I}^{(i)}(\mathbf{x}_1) \ldots  {\cal I}^{(i)}(\mathbf{x}_n) \right \rangle,
\end{equation}
where ${\cal I}^{(i)}({\bf x})$ is the {\it indicator function} for phase $i=1, 2$, defined as 
\begin{align}
  {\cal I}^{(i)}({\bf x})\equiv \begin{cases} 1, &{\bf x} \mathrm{~in~phase~}i \\ 0, &\mathrm{otherwise} \end{cases},
\end{align}
where $n=1,2,3,\ldots$, \cite{Chi13}
and angular brackets denote an ensemble average.  The function $S_n^{(i)}({\bf x}_1, \ldots, {\bf x}_n )$ 
also has a probabilistic interpretation, namely, it is the
probability that the $n$ positions ${\bf x}_1, \ldots, {\bf x}_n$ all lie in phase $i$.
For statistically homogeneous media, $S_{n}^{(i)} \left( \mathbf{x}_1, ..., \mathbf{x}_n \right)$
is translationally invariant and hence depends only on the relative displacements of the points.

The {\it autocovariance} function $\chi_{_V}({\bf r})$, which is directly related to the two-point function $S_2^{(i)}({\bf r})$
and plays a central role in this paper, is defined by
\begin{equation}
\chi_{_V}({\bf r}) \equiv S_2^{(1)}({\bf r})-\phi_1^2=S_2^{(2)}({\bf r})-\phi_2^2,
\label{covariance}
\end{equation}
where ${\bf r}\equiv {\bf x}_2 -{\bf x}_1$.
Here, we have assumed statistical homogeneity.
At the extreme limits of its argument, $\chi_{_V}({\bf r})$ has the following asymptotic behavior:
$\chi_{_V}({\bf r}=0)=\phi_1\phi_2$ and  $\lim_{|{\bf r}| \rightarrow \infty} \chi_{_V}({\bf r})=0$ if
the medium possesses no long-range order.
If the medium is statistically homogeneous and isotropic, then the  autocovariance
function ${\chi_{_V}}({\bf r})$ depends only on the magnitude of its argument $r=|\bf r|$,
and hence is a radial function. In such instances, its slope at the origin is directly related
to the {\it specific surface} $s$, which is the interface area per unit volume. In particular, the well-known
three-dimensional asymptotic result \cite{De57} is easily obtained  in any space
dimension $d$:
\begin{equation}
\chi_{_V}({\bf r})= \phi_1\phi_2 - \kappa(d) s \;|{\bf r}| + {\cal O}(|{\bf r}|^2),
\label{specific}
\end{equation}
where
\begin{equation}
\kappa(d)= \frac{\Gamma(d/2)}{2\sqrt{\pi} \Gamma((d+1)/2)}.
\label{kappa}
\end{equation}

The nonnegative spectral density ${\tilde \chi}_{_V}({\bf k})$, which can be obtained from  scattering experiments \cite{De49,De57},
is the Fourier transform of a well-defined integrable autocovariance function $\chi_{_V}({\bf r})$ \cite{Ko03, Gel64} at wavevector $\bf k$, i.e.,
\begin{equation}
{\tilde \chi}_{_V}({\bf k}) = \int_{\mathbb{R}^d} \chi_{_V}({\bf r}) e^{-i{\bf k \cdot r}} {\rm d} {\bf r} \ge 0, \qquad \mbox{for all} \; {\bf k}.
\label{spectral}
\end{equation}
For a general statistically homogeneous two-phase medium, the spectral density must obey the following sum rule \cite{To20}:
\begin{equation}
\frac{1}{(2\pi)^d}\int_{\mathbb{R}^d} {\tilde \chi}_{_V}({\bf k})\, d{\bf k}= \chi_{_V}({\bf r}=0)=\phi_1\phi_2.
\label{sum}
\end{equation}
 For statistically isotropic media, the spectral density only depends
on the wavenumber $k=|{\bf k}|$ and, as a consequence of  (\ref{specific}), its decay in the large-$k$ limit is controlled
by the exact following power-law form:
\begin{equation}
{\tilde \chi}_{_V}({\bf k}) \sim \frac{\gamma(d)\,s}{k^{d+1}}, \qquad k \rightarrow \infty,
\label{decay}
\end{equation}
where
\begin{equation}
\gamma(d)=2^d\,\pi^{(d-1)/2} \,\Gamma((d+1)/2).
\label{gamma}
\end{equation}

In the case of a packing of identical particles (nonoverlapping particles) ${\cal P}$ of volume $v_1({\cal P})$
at number density $\rho$,
the spectral density $\tilde{\chi}_{_V}(\textbf{k})$ is directly related to the
structure factor $S({\bf k})$ of the particle centroids \cite{To02a,To16b,To18a}:
\begin{align}
  \tilde{\chi}_{_V}(\textbf{k})= \phi_2\, \frac{|\tilde{m}({\bf k};{\cal P})|^2}{v_1({\cal P})}\, S(\textbf{k}),
  \label{chi-packing}
\end{align}
where  $\tilde{m}({\bf k};{\cal P})$, called the {\it form factor}, is the Fourier transform of the particle indicator function
so that $\tilde{m}(0;{\cal P})=v_1({\cal P})$, and
\begin{equation}
\phi_2=\rho v_1({\cal P})
\end{equation}
is the packing fraction, i.e., the fraction of space covered by the identical nonoverlapping particles.
For example, in the case of identical $d$-dimensional spheres of radius $a$, the form factor
is given by
\begin{align}
  \tilde{m}(k;a) = \left( \frac{2\pi a}{k} \right)^{d/2}\, J_{d/2}(ka).
  \label{indicator}
\end{align}
For any such sphere packing, the specific surface is given by
\begin{equation}
s= \frac{\phi_2\,d}{a}.
\label{s-sphere}
\end{equation}

Stealthy hyperuniform  media are a subclass of hyperuniform media that belong to class I. They are defined to possess
zero-scattering intensity for a set of wavevectors around the origin \cite{To16b}, i.e.,
\begin{equation}
{\tilde \chi}_{_V}({\bf k})=0 \qquad \mbox{for}\; 0 \le |{\bf k}| \le K.
\label{stealth}
\end{equation}
Examples of such media are periodic packings of spheres, unusual disordered sphere packings derived from stealthy point patterns, as well as specially designed stealthy hyperuniform dispersions \cite{To16b,Zh16b,Ch18a}.

\subsection{Large-$R$ Asymptotic Analysis of the Variance}
\label{asy}

For a large class  of statistically homogeneous two-phase media in $\R^d$, the large-$R$ asymptotic expansion
of the local volume-fraction variance $\vv{R}$ is given by \cite{Za09}:
\begin{equation}\label{large-R-sig}
\vv{R} = \bar{A}_{_V}  \qty(\frac{D}{R})^d + \bar{B}_{_V} \qty(\frac{D}{R})^{d+1} + o\qty(\frac{D}{R})^{d+1},
\end{equation}
where $\bar{A}_{_V}$ and $\bar{B}_{_V}$ are dimensionless asymptotic coefficients of powers $R^{-d}$ and $R^{-(d+1)}$, respectively, given by 
\begin{align}
 \bar{A}_{_V}  =& \frac{1}{\fn{v_1}{D}} \int_{\mathbb{R}^d} \fn{\chi_{_V}}{\vect{r}}\dd{\vect{r}} =\frac{{\tilde \chi}_{_V}({\bf k}=0)}{v_1(D)},\label{A}\\
 \bar{B}_{_V}  =& -\frac{\fn{c}{d}}{2D\fn{v_1}{D}} \int_{\mathbb{R}^d} \fn{\chi_{_V}}{\vect{r}}\abs{\vect{r}}\dd{\vect{r}},\label{B}
\end{align}
$\fn{c}{d}\equiv 2\fn{\Gamma}{1+d/2}/\qty[\pi^{1/2} \fn{\Gamma}{(d+1)/2}]$,  $D$ is a characteristic microscopic length scale of the medium,
and $o\qty(D/R)^{d+1}$ represents terms of order higher than $(D/R)^{d+1}$. For typical nonhyperuniform media,
$\bar{A}_V$ is positive [cf. (\ref{sigma-nonhyper})]. When $\bar{A}_V=0$, $\bar{B}_V$ must be positive,
implying that the medium is hyperuniform of class I [cf. (\ref{sigma-hyper})].
It is noteworthy that, unlike $\vv{R}$, the coefficient $\bar{B}_{_V}$ depends on the choice of the length scale $D$. 
\ref{asy-gen} provides a more general
asymptotic expansion of the local volume-fraction variance. Finally, we note
that for any packing of identical particles, formulas (\ref{chi-packing})
and (\ref{A}) yield the leading-order asymptotic coefficient to be generally
given by
\begin{equation}
{\bar A}_{_V}= \phi_2 S(0),
\label{A-packing}
\end{equation}
which was first derived in Ref. \cite{Za09}.


\section{Fourier-Space Representation of the Asymptotic Coefficient $\bar{B}_V$}
\label{Fourier}

Here we derive a Fourier-space representation of the asymptotic coefficient $\bar{B}_{_V}$
for any homogeneous two-phase system, whether hyperuniform or not, provided that the spectral density
meets certain mild conditions. This representation will be especially useful when the scattering
intensity is available experimentally or if the spectral density
is known analytically. Specifically, the coefficient $\bar{B}_V$ can be expressed as follows:
\begin{equation}
\bar{B}_{_V}=\frac{\Gamma(1+d/2)d}{ \pi^{(d+2)/2} D^{d+1}} \int_0^\infty \frac{{\tilde \chi}_{_V}(k)-{\tilde \chi}_{_V}(0)}{k^2} \, dk,
\label{Fourier-B}
\end{equation}
where ${\tilde \chi}_{_V}(0) \equiv \lim_{|{\bf k}|\to 0} {\tilde \chi}_{_V}({\bf k })$.
Thus, this Fourier-space representation of the coefficient $\bar{B}_{_V}$ is bounded provided that the difference $[ {\tilde \chi}_{_V}(k)-{\tilde \chi}_{_V}(0)]$ tends to zero in the limit $k \to 0$ faster than linear in $k$.
This condition will always be met by any spectral density that is analytic at the origin, since $[ {\tilde \chi}_{_V}(k)-{\tilde \chi}_{_V}(0)]$ must vanish
at least as fast as quadratically in $k$ as $k \to 0$. In this paper, we will often use formula (\ref{Fourier-B}) to determine $\bar{B}_{_V}$, either analytically or numerically.

To prove the formula (\ref{Fourier-B}), we begin by  using  the identity \cite{To03a}
\begin{equation}
\frac{1}{(2\pi)^d} \int_{\mathbb{R}^d} {\tilde \alpha}_2(k; R) d{\bf k}=1,
\label{identity}
\end{equation}
in relation (\ref{sig-Fourier}) to yield 
\begin{equation}
\sigma^2_{_V}(R)=  \frac{{\tilde \chi}_{_V}(0)}{v_1(R)} +\frac{1}{v_1(R) (2\pi)^d} \int_{\mathbb{R}^d} [{\tilde \chi}_{_V}({\bf k})-{\tilde \chi}_{_V}(0)] 
{\tilde \alpha}_2(k; R) d{\bf k}.
\label{var-2}
\end{equation}
Since ${\tilde \alpha}_2(k; R)$ is a radial function, depending only on the magnitude
of the wavevector, we can carry out the angular integration in the integral in (\ref{var-2}), yielding 
\begin{equation}
\sigma^2_{_V}(R) =  \frac{{\tilde \chi}_{_V}(0)}{v_1(R)} +\frac{d}{R^d (2\pi)^d} \int_0^\infty k^{d-1} [ {\tilde \chi}_{_V}(k)-{\tilde \chi}_{_V}(0)] 
{\tilde \alpha}_2(k; R) dk,
\label{var-3}
\end{equation}
where the radial function ${\tilde \chi}_{_V}(k)$ is given by
\begin{equation}
{\tilde \chi}_{_V}(k) = \frac{1}{\Omega} \int_{\Omega} {\tilde \chi}_{_V}({\bf k}) d\Omega,
\label{spec-radial}
\end{equation}
where $d\Omega$ is the differential solid angle and $\Omega = \frac{d \pi^{d/2}}{\Gamma(1+d/2)}$
is the total solid angle contained in a $d$-dimensional sphere. For large $R$,
\begin{equation}
{\tilde \alpha}_2(k;R) \sim 2^{d+1} \pi^{d/2-1} \Gamma(1+d/2)\frac{ \cos^2\left[kR -\frac{d+1}{4}\right]}{R k^{d+1}}.
\label{asym}
\end{equation}
Combination of (\ref{var-3}) and (\ref{asym}) yields the following large-$R$ asymptotic expansion:
\begin{equation}
\sigma^2_{_V}(R) \sim  \frac{{\tilde \chi}_{_V}({\bf k}\to 0)}{v_1(R)} +\frac{2 \Gamma(1+d/2)\,d}{R^{d+1}\pi^{(d+2)/2} } \int_0^\infty \frac{[ {\tilde \chi}_{_V}(k)-
{\tilde \chi}_{_V}(0)]}{k^2}  \cos^2\left[kR -\frac{d+1}{4}\right] dk +{\cal O}\left( \frac{1}{R^{d+3}}\right).
\label{var-4}
\end{equation}
Using the identity
\begin{equation}
\lim_{L \to \infty} \frac{1}{L} \int_0^L \cos^2\left[kR -\frac{d+1}{4}\right]\, dR=\frac{1}{2}
\end{equation}
and  (\ref{var-4}), we obtain
\begin{equation}
\sigma^2_{_V}(R)\sim  \frac{{\tilde \chi}_{_V}(0)}{v_1(R)} +\frac{\Gamma(1+d/2)\,d}{R^{d+1}\pi^{(d+2)/2} } \int_0^\infty 
\frac{[ {\tilde \chi}_{_V}(k)-{\tilde \chi}_{_V}(0)]}{k^2} 
 dk +{\cal O}\left( \frac{1}{R^{d+3}}\right).
\label{var-5}
\end{equation}
Comparing (\ref{var-5}) to (\ref{large-R-sig}) yields the desired Fourier-space representation of the
surface-area coefficient $\bar{B}_{_V}$ given by (\ref{Fourier-B}). Finally, we observe that if the coefficient ${\bar B}_{_V}$ is identically zero, relation (\ref{Fourier-B})
leads to the integral condition
\begin{equation}
\int_0^\infty \frac{\tilde{\chi}_{_V}(k)-\tilde{\chi}_{_V}(0)}{k^2} \, dk=0,
\end{equation}
which is the analog of the Fourier-space sum rule for {\it hyposurficial} point configurations \cite{To03a}.

\section{Two-Phase Media Models}
\label{models}

\subsection{Antihyperuniform Media}

We consider   the following autocovariance function corresponding to a model of  antihyperuniform media in three dimensions devised by Torquato \cite{To21b}:
\begin{equation}
\frac{\chi_{_V}(r)}{\phi_1\phi_2} =\frac{1}{1+2(r/a) +(r/a)^2},
\label{auto-anti}
\end{equation}
whose specific surface is given by
\begin{equation}
    s=\frac{8\phi_1\phi_2}{a}.
\end{equation}
This monotonic functional form meets all of the known necessary realizability conditions on a valid autocovariance function \cite{To16b}. The corresponding spectral density is given by
\begin{equation}
{\tilde \chi}_{_V}(k)= \frac{4\pi a^2}{ka}  \Big\{ \mbox{Ci}(ka)[ ka \cos(ka)+\sin(ka)]+\mbox{Ssi}(ka)[ka\sin(ka)-\cos(ka)\Big\} ,
\label{spec-anti}
\end{equation}
where $\mbox{Ci}(x)\equiv \int_0^x dt \cos(t)/t$ is the cosine integral, $\mbox{Ssi}(x) \equiv \mbox{Si}(x) -\pi/2$ is the shifted sine integral
and $\mbox{Si}(x) \equiv \int_0^x dt \sin(t)/t$ is the sine integral.
We see that ${\tilde \chi}_{_V}(k)\sim2\pi^2/k$ in the limit $k\to 0$, which is consistent
with the power-law decay
 $1/r^2$ of $\chi_{_V}(r)$ in the limit $r \to \infty$.

\subsection{Debye Random Media}

 Debye {\it et al.} \cite{De57} hypothesized
that the following autocovariance function characterizes  isotropic random media
in which the  phases form domains of ``random shape and size:"
\begin{equation}
\chi_{_V}(r)=\phi_1\phi_2 \exp(-r/a),
\label{Debye}
\end{equation}
where $a$ is a characteristic length scale. 
The Taylor expansion of (\ref{Debye}) about $r=0$ and comparison to (\ref{specific}) reveals
that the specific surface $s$ of a Debye random medium in any space dimension is given by
\begin{equation}
s= \frac{\phi_1\,\phi_2}{\kappa(d)\,a},
\end{equation}
The spectral density for Debye random media in any space dimension is given by \cite{To20}
\begin{equation}
{\tilde \chi}_{_V}(k) = \frac{ \phi_1\phi_2\, c_d\, a^d}{[1+(ka)^2]^{(d+1)/2}},
\end{equation}
where $c_d=2^d \pi^{(d-1)/2} \Gamma((d+1)/2)$.

\subsection{Overlapping Spheres}\label{sec:overlapping-spheres}

    The model of overlapping spheres or fully-penetrable-sphere model refers to
 an uncorrelated (Poisson) distribution of spheres of radius $a$ throughout a matrix \cite{To02a}.
For such nonhyperuniform models at number density $\rho$ in $d$-dimensional Euclidean space $\mathbb{R}^d$, the autocovariance function is known analytically \cite{To02a}:
\begin{equation}\label{eq:autoco_OS}
\fn{\chi_{_V}}{r} = 
\exp\qty(-\rho   v_2(r;a))-{\phi_1}^2,
\end{equation}
where $\phi_1 = \exp(-\rho\fn{v_1}{a} )$ is the volume fraction of the matrix phase (phase 1), $v_1(a)$ is given by (\ref{v1}),
and  $\fn{v_2}{r;a}$ represents the union volume of two spheres whose centers are separated by a distance  $r$.
In two and three dimensions, the latter is explicitly given respectively by
\begin{align}
\frac{\fn{v_2}{r;a}}{\fn{v_1}{a}} = 
\begin{cases} 
2 \fn{\Theta}{x-1}+\qty(1+x) \fn{\Theta}{1-x} , & d=1\\
 2 \fn{\Theta}{x-1}+ \frac{2}{\pi} \Big[\pi + x \qty(1-x^2)^{1/2} -\fn{\cos^{-1}}{x} \Big] \fn{\Theta}{1-x}  
, & d=2 \\
2 \fn{\Theta}{x-1}+\qty(1+\frac{3x}{2}-\frac{x^3}{2}) \fn{\Theta}{1-x} , & d=3
\end{cases}
\end{align}
where $x\equiv r/2a$, and $\Theta(x)$ (equal to 1 for $x>0$ and zero otherwise) is the Heaviside step function.
The specific surface $s$ in any space dimension is given by \cite{To02a}
\begin{equation}
s=\frac{\eta\phi_1 \,d}{a},
\end{equation}
where $\eta\equiv\rho v_1(a)$. For $d=1$, the spectral density can be expressed in the following closed-form:
\begin{equation}
 \tilde{\chi}_{_V}(k) = \frac{2\phi_1\eta}{k((2ak)^2+\eta^2)}\left[2ak(1 - \phi_1\cos(2ak)) + \phi_1\eta\sin(2ak)\right].
\end{equation}

\subsection{Random Checkerboard}

The random checkerboard in $d$ dimensions is generated by tessellating space
into identical hypercubic cells of side length $a$ and randomly designating a cell as phase 1 or 2 with probability $\phi_1$ or $\phi_2$, respectively.
The angular-averaged autocovariance takes the form \cite{To02a}
\begin{equation}
\chi_{_V}(r) =W(r)\phi_1\phi_2,
\end{equation}
where $W(r)$ is a radial function with support in the interval $[0,\sqrt{d}\,a]$.
For example, for $d=1$,
\begin{align*}
W(r)=(1-x)\Theta(1-x),
\end{align*}
 and for $d=2$,
\begin{align}
W(r)=
\begin{cases}
1+\frac{x^2-4x}{\pi}, & 0 \le x \le 1\\
1- \frac{2+x^2}{\pi} +\frac{4}{\pi} \Big[  \qty(x^2-1)^{1/2} -\fn{\cos^{-1}}{1/x} \Big], & 1 \le x \le \sqrt{2} \\
0 , &  x \ge \sqrt{2},
\end{cases}
\end{align}
where $x=r/a$. The explicit expression for $W(r)$ for $d=3$ is given in Ref. \cite{To02a}.
The specific surface $s$ in any space dimension is given by \cite{To02a}
\begin{equation}
s=\frac{2d\,\phi_1\phi_2}{a}.
\end{equation}
For $d=1$, the spectral density can expressed in the following closed-form:
\begin{equation}
 \tilde{\chi}_{_V}(k) = \phi_1\phi_2 a \frac{\sin^2(ka/2)}{(ka/2)^2}.
\end{equation}

\subsection{Equilibrium Packings}

We also examine equilibrium (Gibbs) ensembles of identical hard spheres of radius $a$ at packing fraction $\phi_2$ \cite{Ha86,To18b}. In particular, we consider such disordered packings along the stable disordered fluid branch in the phase diagram \cite{To02a,To18b}.
All such states are nonhyperuniform. In the case of 1D equilibrium hard rods, pair statistics
are known exactly \cite{Pe64}. In particular, using the exact solution of the 
direct correlation function \cite{Ze27,Pe64} and the Ornstein-Zernike integral equation,
we can express the exact structure factor as
    \begin{align*}
        S(k) = \left[ 1 - \frac{2\phi_2\left\{ \phi_2\left[\cos(2ak)-1\right] +2ak\sin(2ak)(\phi_2-1) \right\}}{(1-\phi_2)^2(2ak)^2} \right]^{-1}. \\
    \end{align*}
For $d=3$, we utilize the Percus-Yevick approximation of the structure factor $S(k)$ \cite{Ha86}:
\begin{align}
\fn{S}{k} = 
&\Big(1-\rho \frac{16 \pi  a^3 }{q^6} 
\Big\{\big[24 a_1 \phi_2 - 12 (a_1 + 2 a_2) \phi_2 q^2 
    \nonumber \\
    &\quad+ (12 a_2 \phi_2 + 2 a_1 + a_2\phi_2) q^4] \cos(q) 
    \nonumber \\
&+ [24 a_1 \phi_2 q - 2 (a_1 + 2 a_1 \phi_2 + 12 a_2 \phi_2) q^3\big] \sin(q)
    \nonumber \\
& -24 \phi_2 (a_1 - a_2 q^2) \Big\}\Big)^{-1},
\end{align}
where $q = 2ka$, $a_1 = (1+2\phi_2)^2/(1-\phi_2)^4$, and $a_2 = -(1+\phi_2/2)^2 /(1-\phi_2)^4$.
Using these solutions for the structure factor in conjunction with (\ref{chi-packing}) yields the corresponding spectral density ${\tilde \chi}_{_V}(k)$.
For $d=2$, there is no closed-form approximation for the structure, and so we obtain the spectral density from disk packings generated by the Monte Carlo method \cite{To02a}.

\subsection{Disordered Hyperuniform Media}
\label{disorder-hyper}

We also consider models of hyperuniform two-phase media in $\mathbb{R}^d$ formulated by Torquato \cite{To16b,To21d}
in which the autocovariance function takes the following form:
\begin{equation}
 \frac{\chi_{_V}(r)}{\phi_1\phi_2}=c \,  e^{-r/a}\cos(qr +\theta),
\label{auto-hyper}
\end{equation}
where the parameters  $q$ and $\theta$ are the wavenumber and phase associated with the oscillations of $\chi_{_V}(r)$, respectively,
$a$ is a correlation length, and $c$ is a normalization constant to be chosen so that the right-hand side
of (\ref{auto-hyper}) is unity for $r=0$. 
For $d=1$, the phase is given by
$\theta =\tan^{-1}\left(1/(qa)\right)$, implying that the normalization constant is
$c= [1+ (qa)^2]^{1/2}/(qa)$.  For concreteness, we set $qa=1$, and hence $c=\sqrt{2}$ and $\theta=\pi/4$.
Taking the Fourier transform of (\ref{auto-hyper}) with these parameters  yields
the spectral density to be given by
\begin{equation}
\frac{{\tilde \chi}_{_V}(k)}{\phi_1\phi_2}=  \frac{4\, (ka)^2\, a}{(ka)^4 +4}.
\label{spec1-hyper}
\end{equation}
In higher dimensions, one can take $\theta=0$ and $c=1$. 
The corresponding spectral densities for $d=2$ with $(qa)^2=1$ and $d=3$ with  $(qa)^2=1/3$ are respectively given by
\begin{equation}
\frac{{\tilde \chi}_{_V}(k)}{\phi_1\phi_2}= \frac{ 2\pi (ka)^2 [A(k)+B(k)] + 4\pi [A(k)-B(k)]\, a^2}{[(ka)^4+4] [A^2(k)+B^2(k)] },
\label{spec2-hyper}
\end{equation}
and
\begin{equation}
\frac{{\tilde \chi}_{_V}(k)}{\phi_1\phi_2}= \frac{216\pi \, [3 (ka)^2+8](ka)^2\,a^3}{81 (ka)^8+216 (ka)^6+ 432 (ka)^4+ 384 (ka)^2+256},
\label{spec3-hyper}
\end{equation}
where
\begin{equation}
A(k)=\sqrt{(ka)^2/2+ \sqrt{(ka)^4+4}/2}, \qquad B(k)=A^{-1}(k).
\end{equation}
Note that the specific surface for this system in any dimension $d$ is given by
\begin{equation}
    s = \frac{2\sqrt{\pi}c\phi_1\phi_2\Gamma\left[(1+d)/2\right]\left[\cos(\theta) + qa\sin(\theta) \right]}{a\Gamma[d/2]}.
\end{equation}

\subsection{Stealthy Hyperuniform Media}

We also study ``stealthy" hyperuniform two-phase media, which obey the general functional form
given by (\ref{stealth}), where $K$ is the exclusion sphere radius in Fourier (reciprocal) space. 
One  can create stealthy packings of identical spheres by decorating stealthy point 
configurations, generated via the so-called collective-coordinate optimization 
technique \cite{Uc04b,Zh15a},  by spheres of radius $a$ such that spheres cannot overlap \cite{Zh16}.
Here we utilize a modification of this algorithm by incorporating an additional soft-core repulsive interaction between the points 
to further increase the nearest-neighbor distance so that even higher packing fractions 
can be achieved by a decoration of the points by nonoverlapping spheres \cite{Ki20a,To21a}.
 Disordered stealthy point configurations generated by this optimization procedure are actually classical ground states of systems of 
particles interacting with bounded long-ranged pair potentials.
The corresponding spectral densities in this work are obtained from the numerically generated stealthy packings.

\subsection{Periodic Media}

We consider nonoverlapping particles $\cal P$ on the sites of any  Bravais lattice $\cal L$ in $\mathbb{R}^d$
in which a single particle $\cal{P}$ is placed in a fundamental cell $\mathcal{F}$ of $\cal L$.
One can immediately obtain from \eqref{chi-packing} the specific formulas for the corresponding spectral density as follows:
\begin{align}\label{eq:chik-network-1pt}
\spD{\vect{k}} &= {V_\mathcal{F}}^{-1} \abs{\fn{\tilde{m}}{\vect{k};\cal{P}} }^2 \fn{S_\mathcal{L}}{\vect{k}},  
\end{align}
where $V_\mathcal{F}$ is the volume of $\mathcal{F}$, $\fn{S_\mathcal{L}}{\vect{k}}$ is the structure factor of $\mathcal{L}$ given by \cite{To18a}
\begin{equation}
\fn{S_\mathcal{L}}{\vect{k}} = \frac{(2\pi)^d}{V_\mathcal{F}} \sum_{\vect{q}\in \mathcal{L}^* \setminus\{\vect{0}\}} \fn{\delta}{\vect{k}-\vect{q}},
\end{equation}
where $\mathcal{L}^*$ denotes the reciprocal lattice of $\mathcal{L}$, and $\fn{\delta}{\vect{x}}$ is the Dirac delta function.
Specifically, for $d=1$, we consider rods of phase 2 placed on the sites
of the integer lattice, whose specific surface is given by (\ref{s-sphere}).
For $d=2$, we consider both circular disks and oriented hexagons of side length $a$ placed 
on the sites of the triangular lattice, whose specific surfaces are
given by (\ref{s-sphere}) and $s=4\phi_2/(\sqrt{3}a)$, respectively
\cite{Ki21}.
The form factor $\fn{\tilde{m}}{\vect{k};\cal{P}}$ for the hexagon is obtained using the analysis presented in Ref. \cite{Ki21}.
For $d=3$, we consider spheres on the sites of both the simple cubic (SC) and body-centered cubic (BCC) lattices, whose specific surfaces are given by (\ref{s-sphere}).

\subsection{Representative Microstructure Images}

\begin{figure*}[!htp]
        \subfloat[\label{fig:micro_a}]{\includegraphics[width=0.25\textwidth]{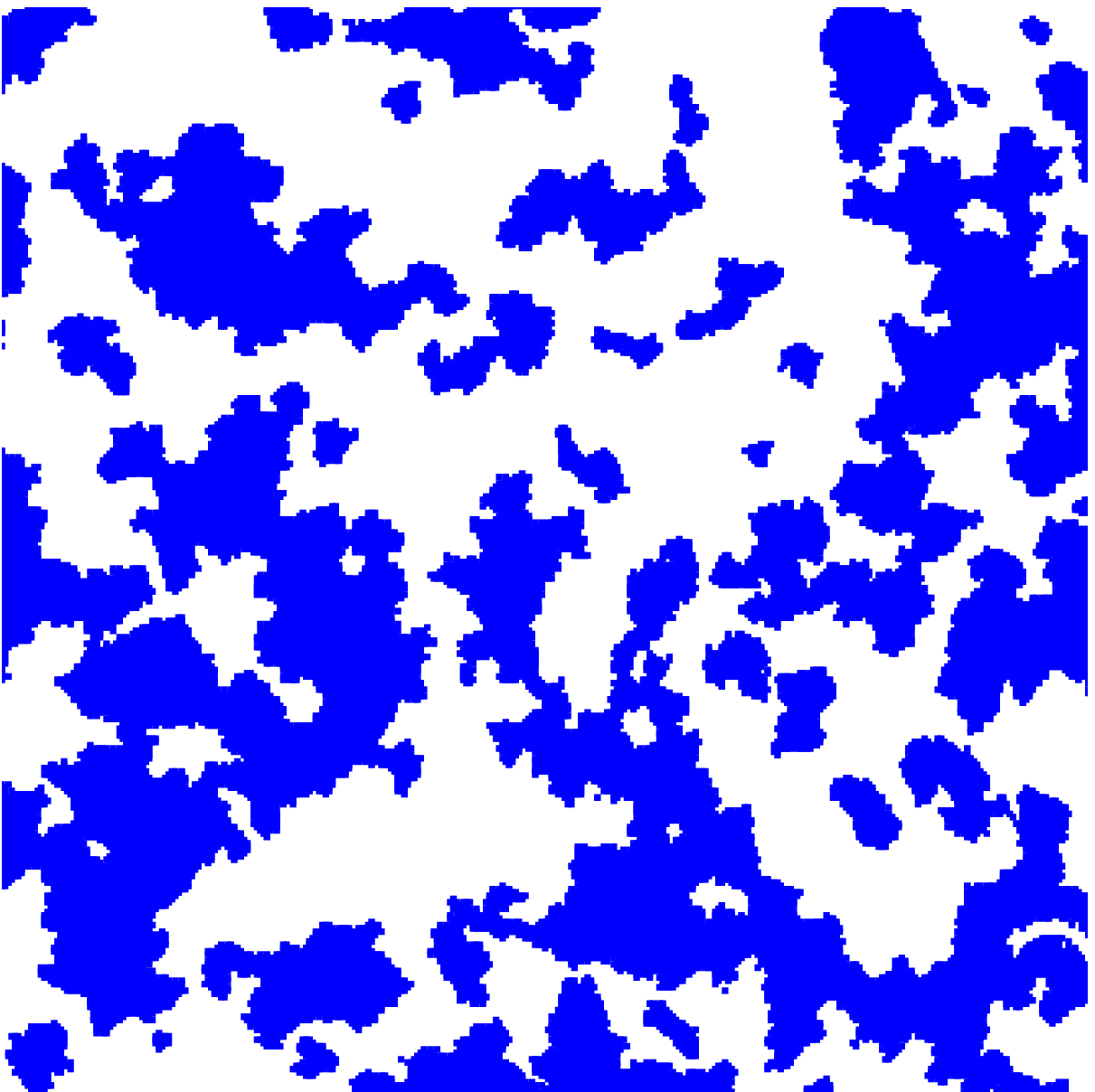}}
    \subfloat[\label{fig:micro_b}]{\includegraphics[width=0.25\textwidth]{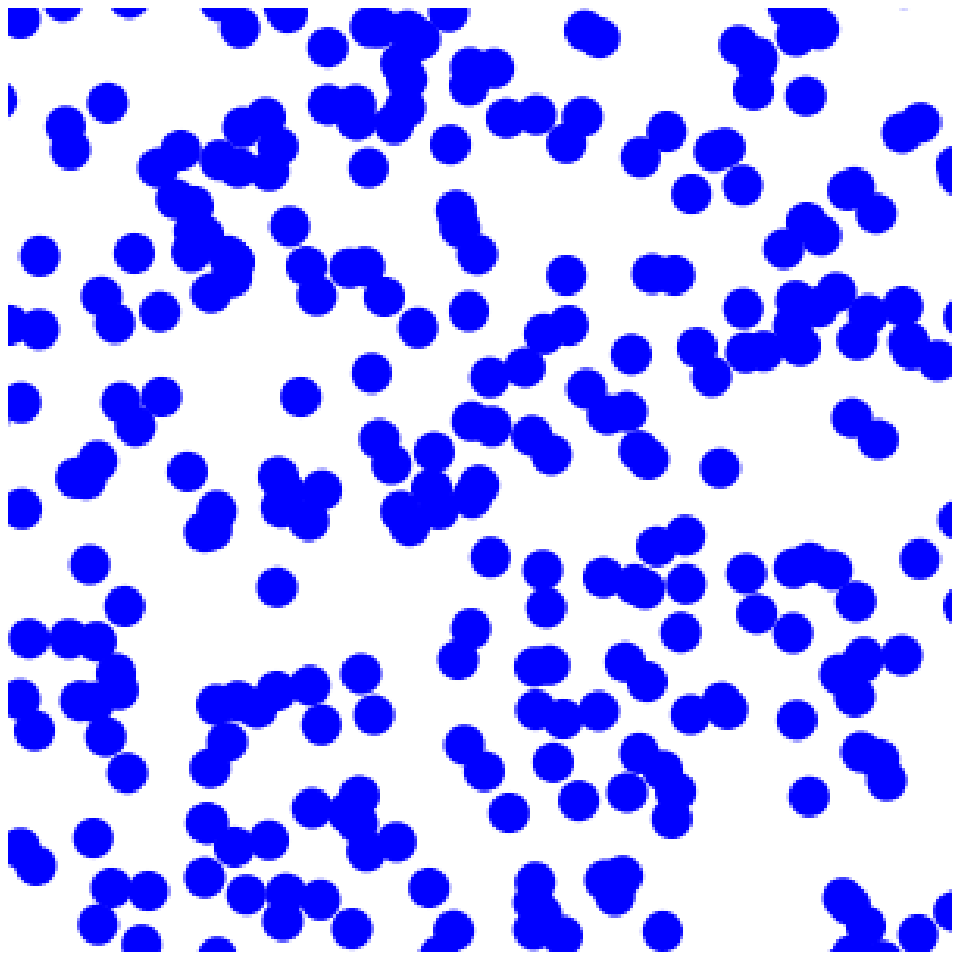}}
    \subfloat[\label{fig:micro_c}]{\includegraphics[width=0.25\textwidth]{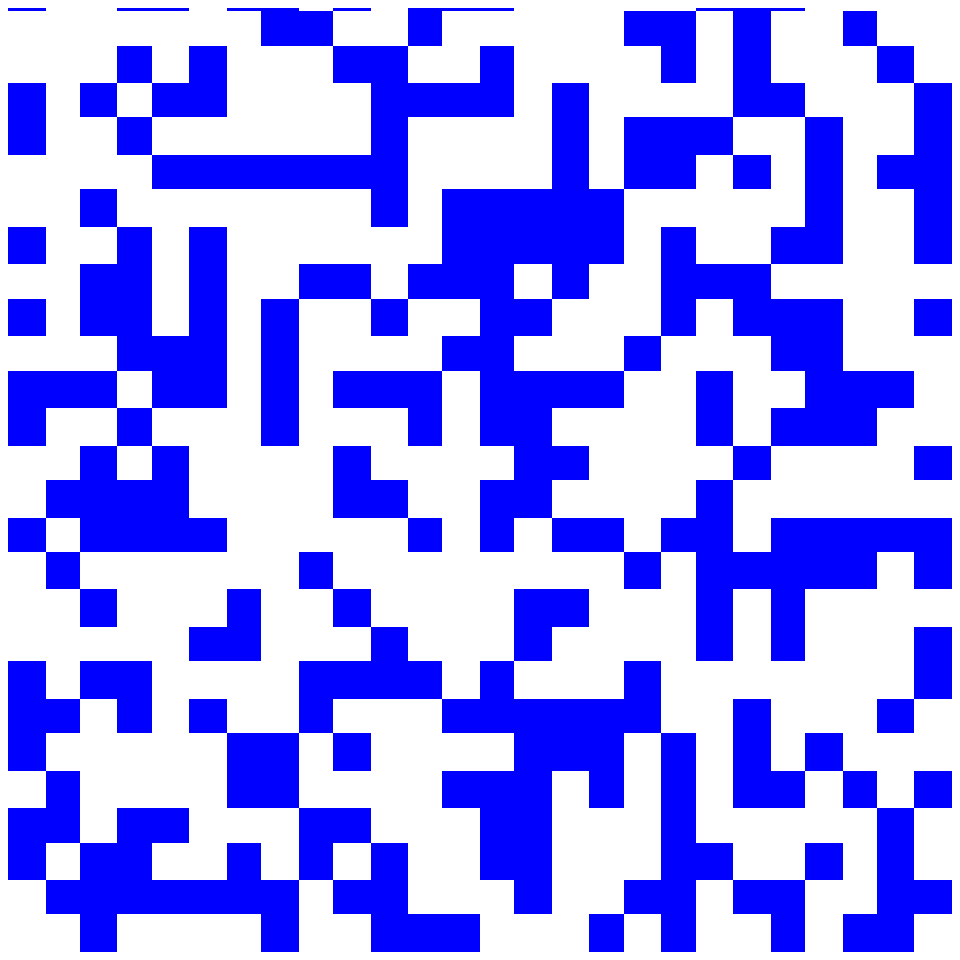}}
    \subfloat[\label{fig:micro_d}]{\includegraphics[width=0.25\textwidth]{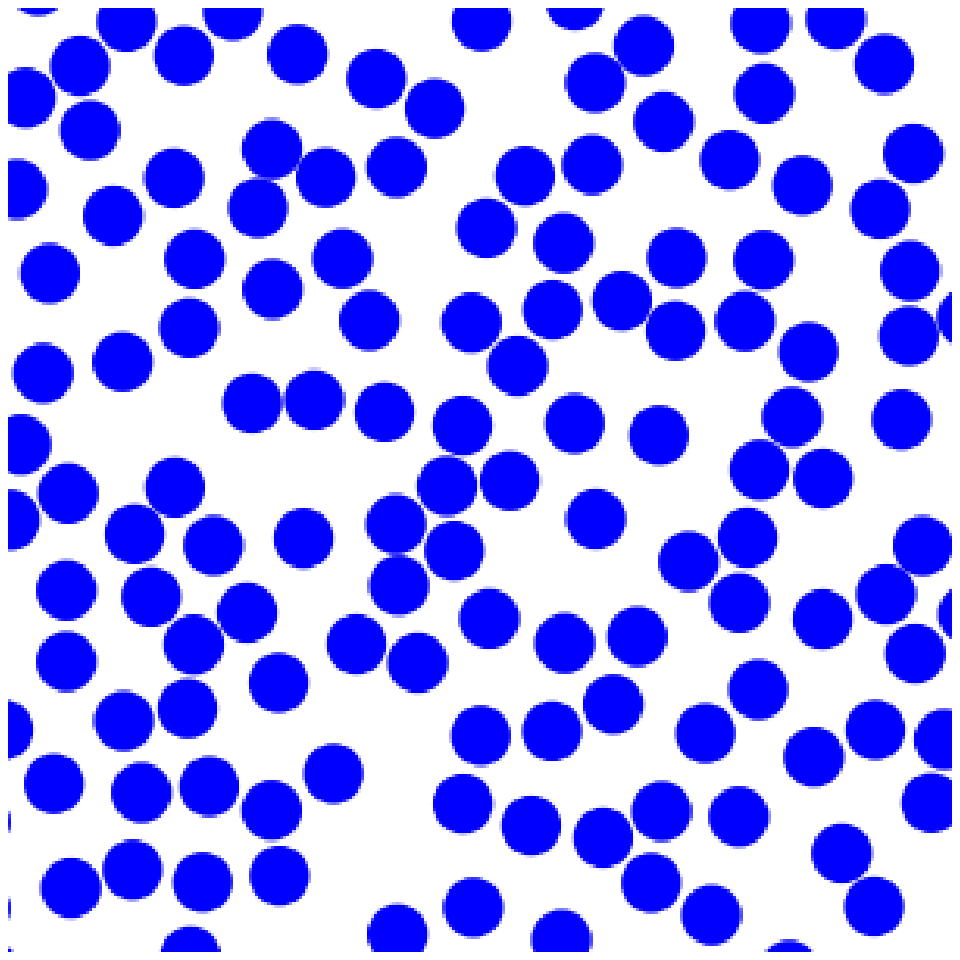}} \\
    \subfloat[\label{fig:micro_e}]{\includegraphics[width=0.25\textwidth]{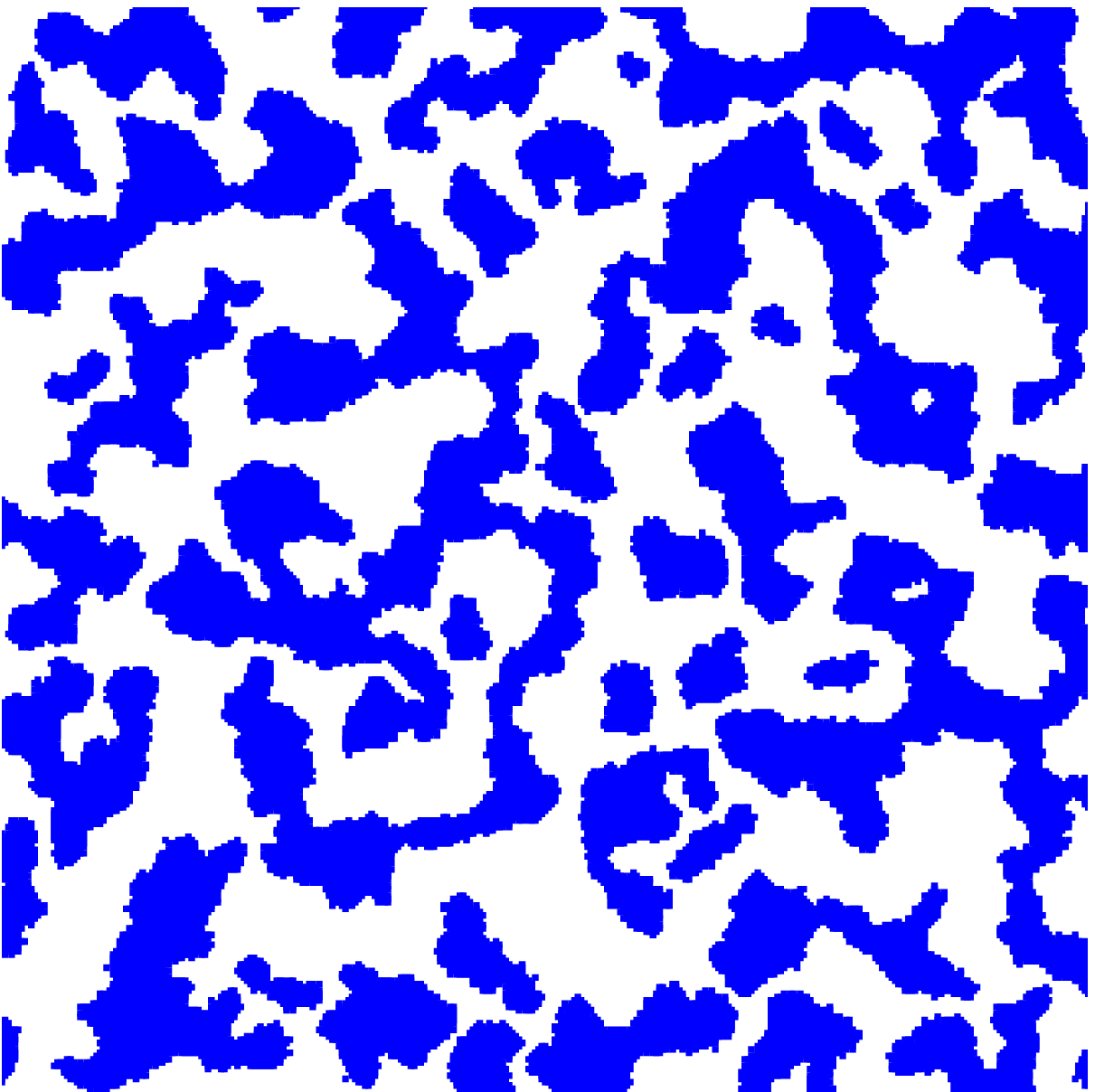}}
    \subfloat[\label{fig:micro_f}]{\includegraphics[width=0.25\textwidth]{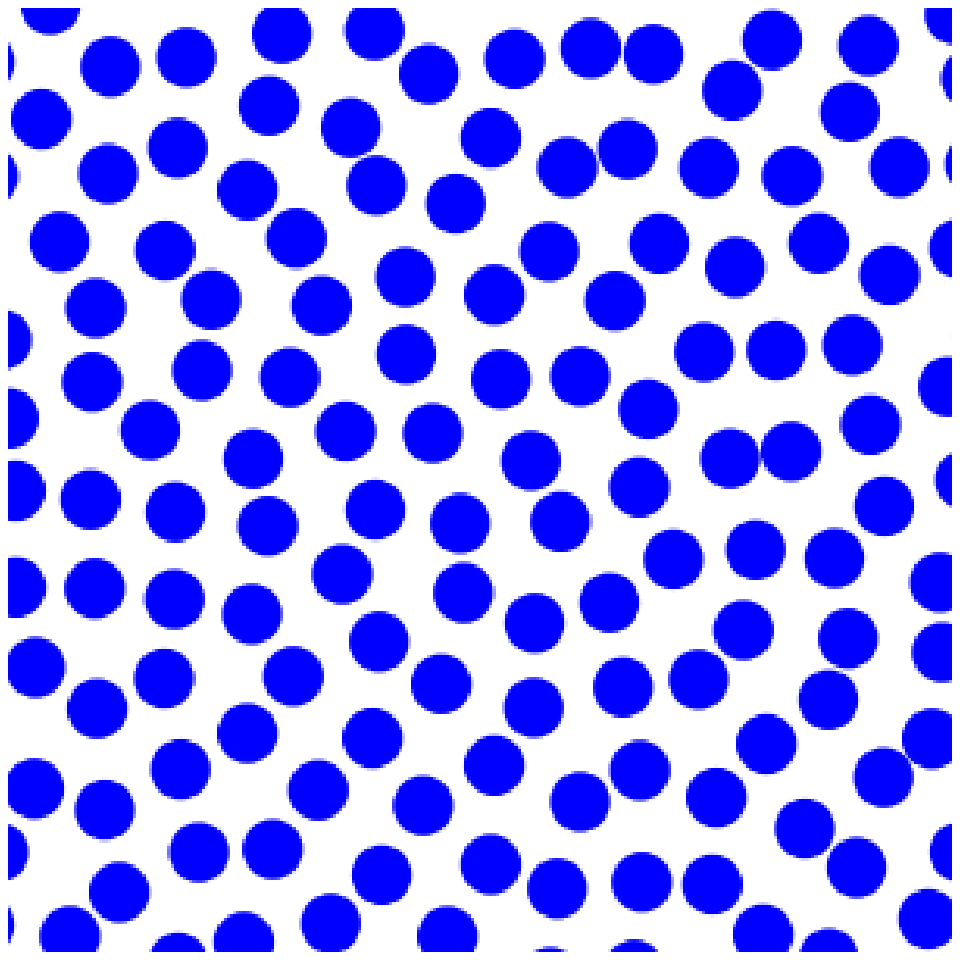}}
    \subfloat[\label{fig:micro_g}]{\includegraphics[width=0.25\textwidth]{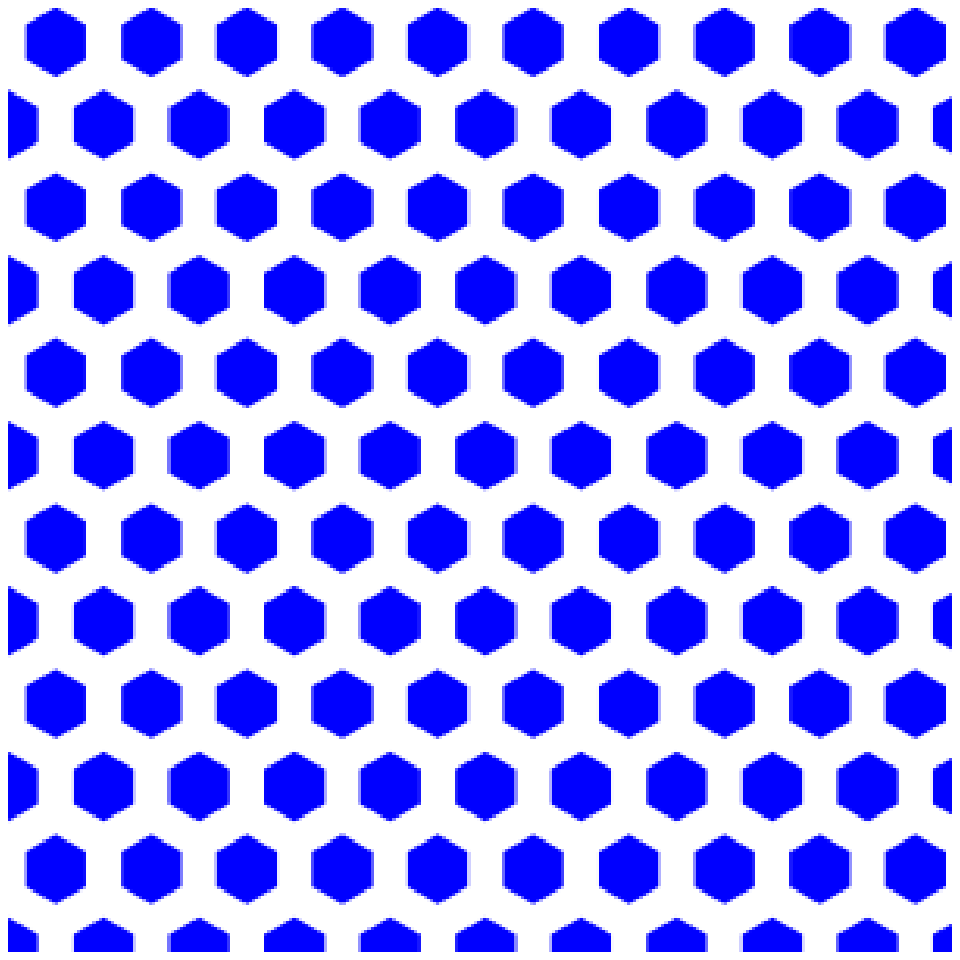}}
    \subfloat[\label{fig:micro_h}]{\includegraphics[width=0.25\textwidth]{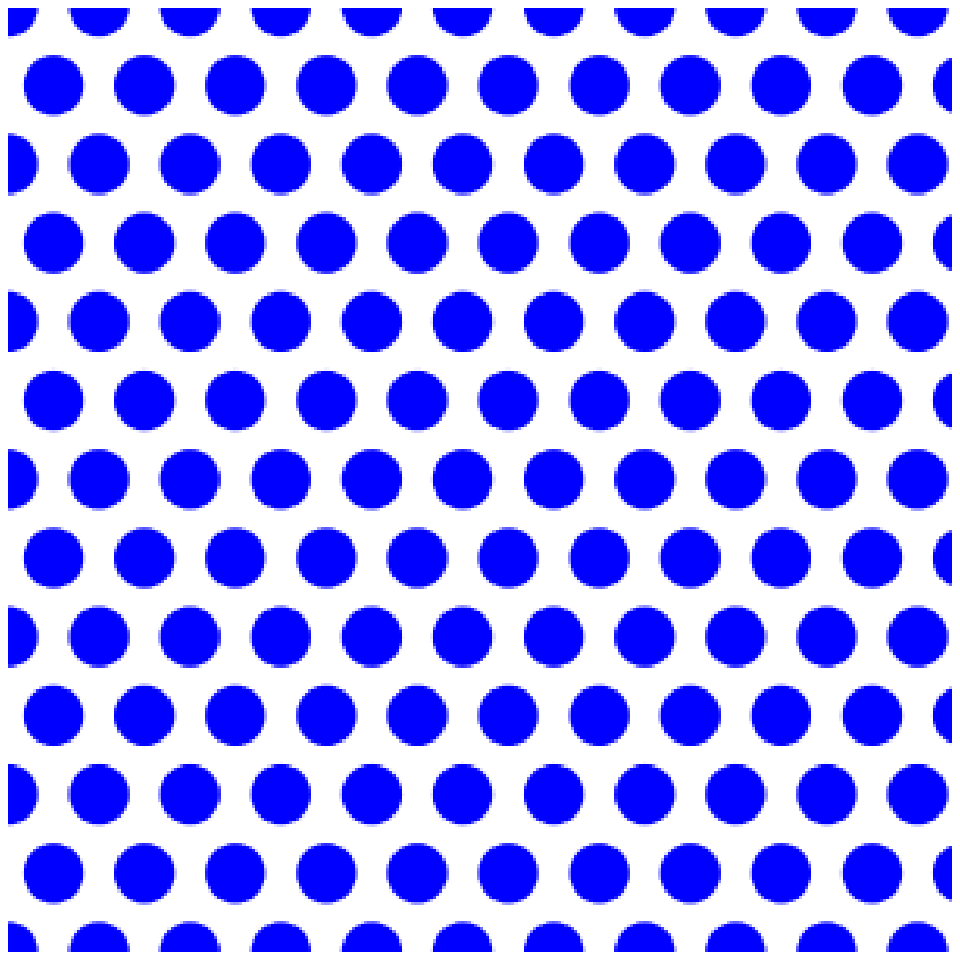}}
\caption{Representative images of each of the eight 2D two-phase models at $\phi_2=0.4$: (a) Debye random media; (b) overlapping circular disks; (c) random checkerboard; (d) equilibrium hard disks; (e) disordered hyperuniform; (f) stealthy hyperuniform disks; (g) hexagons on the triangular lattice; (h) circular disks on the triangular lattice. Here, the white regions
represent phase 1 and the blue regions represent phase 2.}\label{fig:YT2Dstructs}
\end{figure*}

To get a visual sense of the breadth of microstructures considered in this paper that span
from nonhyperuniform to hyperuniform two-phase media  and their corresponding degree of order,
we depict representative images of small portions of the microstructures of each of the eight 2D two-phase models at $\phi_2=0.4$
in Fig. \ref{fig:YT2Dstructs}. It is expected that Debye random media will be the most disordered
at all length scales because they are characterized by phase domains of random shapes
with a wide range of sizes, including a substantial fraction of large
``holes" \cite{To20,Ma20b,Sk21}. We will see that this is indeed the case in Sec. \ref{results-2}, as well as the fact that circular disks on the triangular lattice are the most ordered.

\subsection{Results for the Spectral Densities}
\label{specs}

\begin{figure*}[!htp]
\centering
\includegraphics[width=0.6\textwidth,clip=]{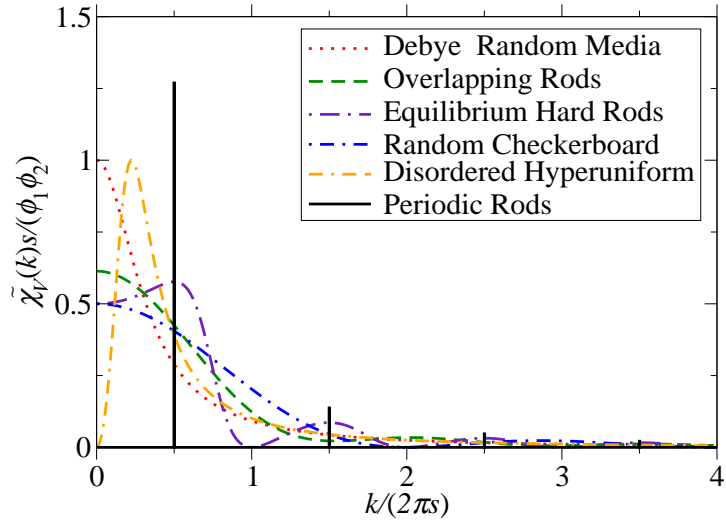}
\caption{Comparison of the  dimensionless spectral densities ${\tilde \chi}_{_V}(k) s/(\phi_1\phi_2)$ versus the dimensionless wavenumber  $k/(2\pi s)$ for 1D models at $\phi_2=0.5$, where $s$ is the specific surface. }\label{1D-spectral}
\end{figure*}

\begin{figure*}[!htp]
\centering
\includegraphics[width=0.6\textwidth,clip=]{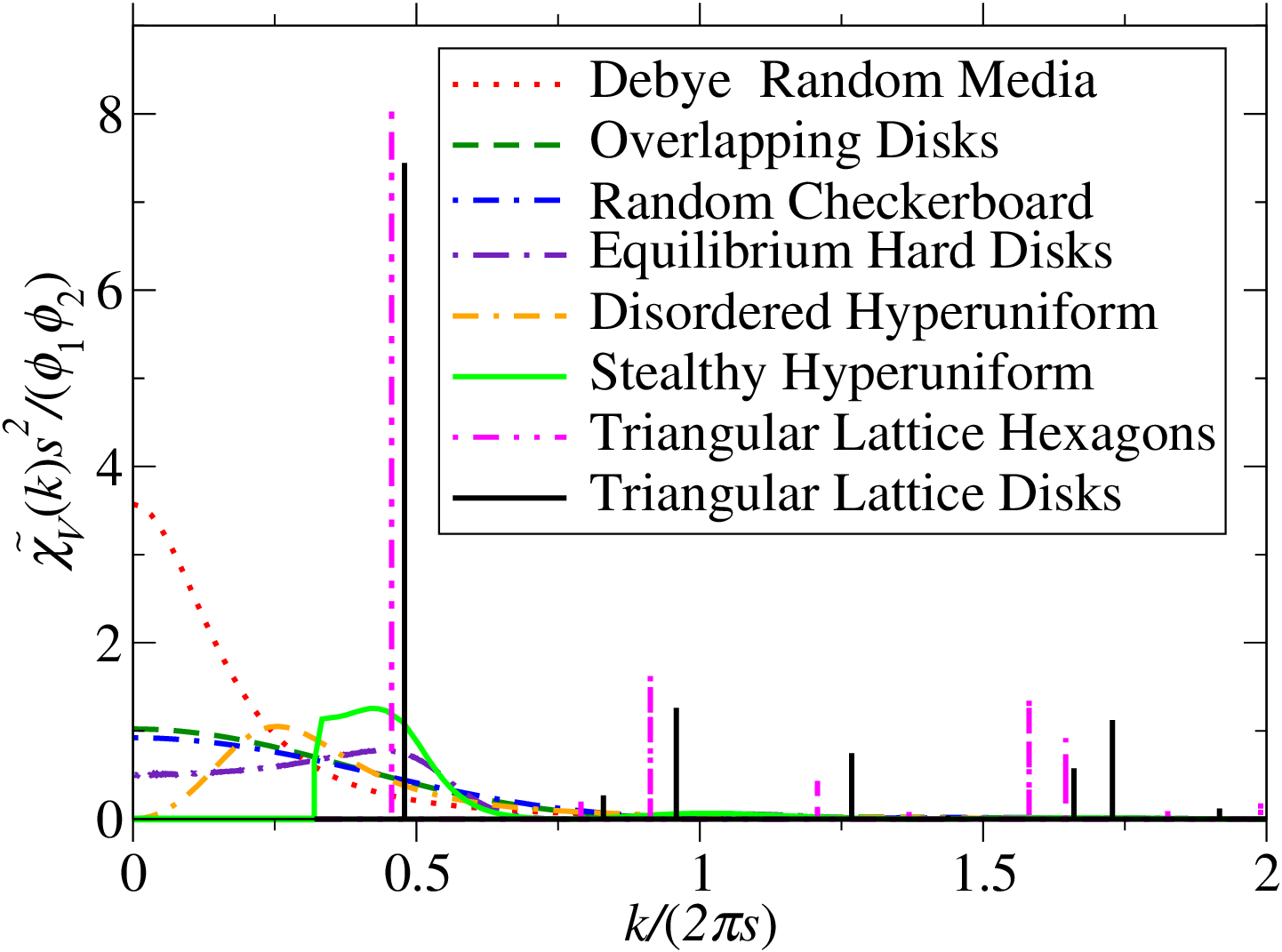}
\caption{Comparison of the dimensionless spectral densities ${\tilde \chi}_{_V}(k) s^2/(\phi_1\phi_2)$ versus the dimensionless wavenumber  $k/(2\pi s)$ for 2D models at $\phi_2=0.4$, where $s$ is the specific surface. In the case
of periodic media, the angular-averaged spectral density is presented. }\label{2D-spectral}
\end{figure*}

\begin{figure*}[!htp]
\centering
\includegraphics[width=0.6\textwidth,clip=]{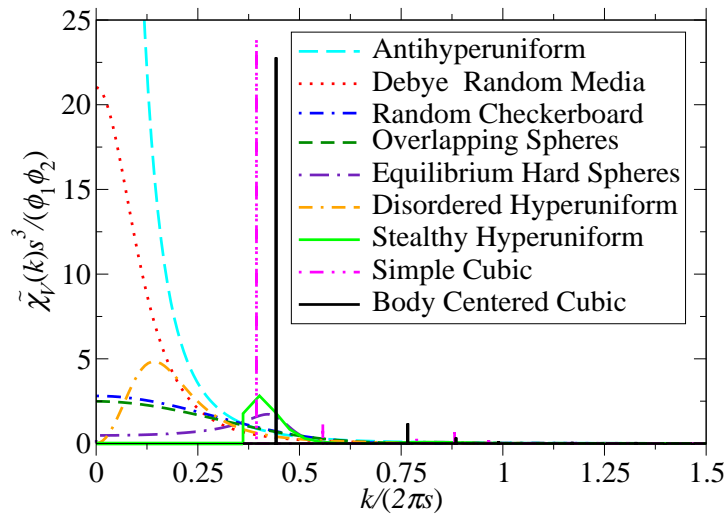}
\caption{Comparison of the dimensionless spectral densities ${\tilde \chi}_{_V}(k) s^3/(\phi_1\phi_2)$ versus the dimensionless wavenumber  $k/(2\pi s)$ for 3D models at $\phi_2=0.38$, where $s$ is the specific surface. In the case
of periodic media, the angular-averaged spectral density is presented.}
\label{3D-spectral}
\end{figure*}

\begin{figure*}[!htp]
\centering
\includegraphics[width=0.6\textwidth,clip=]{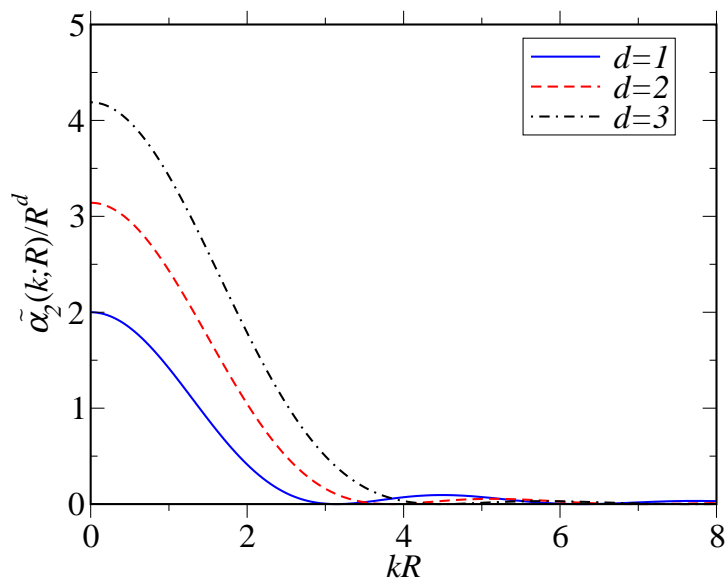}
\caption{Plots of the Fourier transform of the scaled intersection volume of two spherical windows (\ref{alpha}) normalized by $R^d$, $\tilde{\alpha}_2(k;R)/R^d$, as a function of dimensionless wavevector $kR$ for dimensions one, two, and three, as adapted from Ref. \cite{To18a}}\label{ft-intersection}
\end{figure*}

The spectral densities for 1D, 2D and 3D models are depicted in Figs. \ref{1D-spectral}, \ref{2D-spectral}
and \ref{3D-spectral}, respectively. According to formula (\ref{sig-Fourier}) for the local variance $\sigma^2_{_V}(R)$, 
the behavior of the spectral density for small to intermediate wavenumbers determines the magnitude
of the local variance for intermediate to large length scales. In particular, the smaller (larger) 
are the values of ${\tilde \chi}_{_V}(k)$ for such wavenumbers, the smaller (larger) are the values of $\sigma^2_{_V}(R)$ for intermediate to large length scales. More precisely, we see from formula (\ref{sig-Fourier}) that it is the product of the spectral density with function $\tilde{\alpha}_2(k;R)$ [cf. (\ref{alpha})] that determines the behavior of $\sigma_{_V}^2(R)$. We see from the plots of $\tilde{\alpha}_2(k;R)$ for the first three space dimensions shown in Fig. \ref{ft-intersection} that the function (\ref{alpha}) places increasingly heavier weight on the small wavenumber region of the spectral density in integral (\ref{sig-Fourier}) as the dimension increases. Thus, qualitative  changes in the spectral densities for the same
models across dimensions have implications for how their relative order ranking may or may not change across dimensions. 
For example, while the dimensionless spectral density for the 1D random checkerboard is substantially smaller
than that for equilibrium hard rods for a range of wavenumbers near the origin (Fig. \ref{1D-spectral}),
these behaviors for their 2D counterparts are reversed (Fig. \ref{1D-spectral}), which, in turn, 
should reverse their relative rankings, which we will see is indeed the case in Sec. \ref{results-2}.
Across dimensions, periodic media are characterized by Bragg peaks (Dirac delta functions) whose strengths
are proportional to the form factor [cf. (\ref{chi-packing})]. In the 2D and 3D periodic cases,
the structures with the largest first  Bragg peak (i.e., triangular lattice of circles in 2D and BCC lattice of spheres in 3D) should yield the least fluctuations in those dimensions,
which again is verified in Sec. \ref{results-2}. For the same reasons, the stealthy hyperuniform
packings in 2D and 3D should yield the most ordered microstructures among all disordered models.

\section{Local Variance as an Order Metric Across Length Scales}
\label{results-1}

We propose the use of the local volume-fraction variance $\sigma^2_{_V}(R)$ at window radius $R$ as an order
metric for disordered and ordered two-phase media across length scales by tracking it as a function of $R$. Specifically, for any particular value of $R$, the lower the value of $\vv{R}$, the greater the degree of order. 

To extract an integrated measure of local volume-fraction fluctuations for window radii from zero  to some length scale $L$, we consider
the following one-dimensional integral over $\sigma^2_{_V}(R)$:
\begin{equation}
    \Sigma_{_V}(L) = \int_0^L \sigma_{_V}^2(R)dR.
\label{Sigma-L}
\end{equation}
Whenever the integral $\Sigma_{_V}(L)$ converges, i.e., is bounded, in the limit $L\to \infty$, we consider 
\begin{equation}
 \Sigma_{_V}(\infty) \equiv \int_0^\infty \sigma_{_V}^2(R)dR.
\label{Sigma}
\end{equation}
This integral has the following convenient closed-form representation in terms of the angular-averaged spectral density defined by (\ref{spec-radial}):
\begin{equation}
\Sigma_{_V}(\infty) =             
                \frac{  \Gamma(1+d/2) \Gamma(d/2)\,d}
{ 2\pi^{d/2}\Gamma(d+1/2) \Gamma((d+1)/2)}
 \int_{0}^{\infty} k^{d-2} {\tilde \chi}_{_V}(k) dk.
                \label{Sigma-2}
\end{equation}
To prove relation (\ref{Sigma-2}), we substitute formula (\ref{sig-Fourier}) into (\ref{Sigma}) to yield
\begin{eqnarray}
\Sigma_{_V}(\infty) &=&   \frac{1}{(2\pi)^d} \int_{\mathbb{R}^d} {\tilde \chi}_{_V}({\bf k}) d {\bf k}
\int_0^\infty \frac{{\tilde \alpha}_2(k; R)}{v_1(R)} dR \\
&=&   \frac{d}{(2\pi)^d} \int_0^\infty k^{d-1} {\tilde \chi}_{_V}(k) dk
\int_0^\infty \frac{{\tilde \alpha}_2(k; R)}{R^d} dR\\
&=&  \frac{\Gamma(1+d/2)\,d}{\pi^{d/2} } \int_0^\infty k^{d-2} {\tilde \chi}_{_V}(k) dk
\int_0^\infty \frac{J_{d/2}^2(kR)}{ (kR)^d} dR.\label{last}
\end{eqnarray}
Using the identity 
        \begin{equation}
                \int_0^\infty \frac{J_{d/2}(x)^2}{x^d}  \dd{x}= \frac{\fn{\Gamma}{d/2}}{2\fn{\Gamma}{d+1/2} \fn{\Gamma}{(d+1)/2}} 
        \end{equation}
in (\ref{last}) yields (\ref{Sigma-2}). 

For the first three dimensions, Eq. (\ref{Sigma-2}) gives
        \begin{align}
                \Sigma_{_V}(\infty)
                &=
                \begin{cases}
                      \displaystyle  \frac{1}{2} \int_0^{\infty}  \frac{ {\tilde \chi}_{_V}(k)}{k} \dd{k}, & d=1 \\
                        \displaystyle \frac{8}{3\pi^2} \int_0^{\infty} {\tilde \chi}_{_V}(k) \dd{k}, & d=2 \\
                        \displaystyle \frac{3}{10 \pi} \int_0^{\infty}k {\tilde \chi}_{_V}(k) \dd{k}, & d=3.
                \end{cases}
                \label{eq:int-specific}
        \end{align}
Referring to the scaling relations (\ref{spec-scaling}) and (\ref{sigma-hyper}), we see that for $d=1$, 
the integral $\Sigma_{_V}(\infty)$ converges only for hyperuniform media that belong to class I or II. By contrast, it does not converge for $d=1$ for class III hyperuniform media or nonhyperuniform media. 
For typical nonhyperuniform media and hyperuniform media, $\Sigma_{_V}(\infty)$ converges for any $d \ge 2$. For antihyperuniform media,
$\Sigma_{_V}(\infty)$ is nonconvergent if $\alpha$  lies between $-d$ and $2-d$, implying that it is always nonconvergent
for $d=1$ and $d=2$, but for $d=3$, it is nonconvergent only if $\alpha$ lies in the open interval $(-3,-1)$. 
For $d=3$,  $\Sigma_{_V}(\infty)$ is convergent for antihyperuniform media  
if $\alpha$ lies in the open interval $(-1,0)$. Whenever  $\Sigma_{_V}(\infty)$ does not converge, we utilize 
the rate of growth of the integral (\ref{Sigma-L}) with $L$ as the order metric.

\section{Results}
\label{results-2}

In the ensuing description, we present results for the local variance
(as obtained from either (\ref{sig-direct}) or (\ref{sig-Fourier})) and its corresponding integral for
the 1D, 2D and 3D models discussed in Sec. \ref{models}).  However, in order to compare
different models in any particular space dimension, we fix both the volume fraction $\phi_1$
and specific surface $s$. The latter implies that the characteristic microscopic length scale
$D$ is set equal to the inverse of the specific surface, i.e., $D=s^{-1}$. The justification for the
use of the specific surface as a simple means to fix length scales for different media was
provided by Kim and Torquato \cite{Ki21}.

\subsection{1D models}

In what follows, we obtain exact closed-form formulas for $\sigma^2_{_V}(R)$ and  $\Sigma_{_V}(L)$ for five of the six 1D models considered in this 
work, except in the case of equilibrium hard rods, which requires numerical quadrature. For
a particular model, we express formulas in terms of the characteristic microscopic length scales defined
in Sec. \ref{models}. 

For 1D Debye random media, the local variance is given by
\begin{equation}
\sigma^2_{_V}(R)= \phi_1\phi_2  \left(\frac{a}{R}\right)+ \frac{\phi_1\phi_2 [1-\exp(-2R/a)]}{2}\left(\frac{a}{R}\right)^2,
\end{equation}
The large-$R$ asymptotic coefficients in distance units $a$ are given by
\begin{equation}
    \bar{A}_{_V} = \phi_1\phi_2, \qquad \bar{B}_{_V} = -\frac{\phi_1\phi_2}{2}.
\end{equation}
The integral of the variance from $R=0$ to $R=L$ is given by
\begin{align}
    \Sigma_{_V}(L) = \frac{a\phi_1\phi_2}{2}\left[2(\gamma-1) + \frac{a\left(1 - \exp(-2L/a) \right)}{L} - 2\mbox{Ei}\left(-\frac{2L}{a}\right) + \ln(4) + 2\ln\left(\frac{L}{a} \right) \right],
\end{align}
where $\gamma=0.577216...$ is the Euler-Mascheroni constant, and $\mbox{Ei}(x)=-\int_{-x}^{\infty}e^{-t}dt/t$ is the exponential integral.

For the 1D random checkerboard, the local variance is given by
\begin{align}
\sigma^2_{_V}(R)=
\begin{cases}
\phi_1\phi_2 [1-\frac{2}{3}\left(\frac{R}{a}\right)], & R\le a/2 \\
\phi_1\phi_2[  \frac{1}{2}\left(\frac{a}{R}\right) -\frac{1}{12}  \left(\frac{a}{R}\right)^2] , & R\ge a/2.
\end{cases}
\end{align}
The large-$R$ asymptotic coefficients in distance units $a$ are given by
\begin{equation}
    \bar{A}_{_V} = \frac{\phi_1\phi_2}{2}, \qquad \bar{B}_{_V} = -\frac{\phi_1\phi_2}{12}.
\end{equation}
The integral of the variance from $R=0$ to $R=L$ is given by
\begin{align}
    \Sigma_{_V}(L) = \frac{-\phi_1\phi_2}{12aL}\left[4L^2(L-3a) - \Theta\left(L-\frac{a}{2}\right)\left(a^3 - 12aL^2 + 4L^3 + a^2L(3 + \ln(64) + 6\ln\left(\frac{L}{a}\right) \right) \right].
\end{align}

For 1D overlapping rods, the local variance is given by
\begin{align}
\sigma^2_{_V}(R)=
\begin{cases}
-\phi_1^2 + \frac{2\phi_1}{\eta}\left(\frac{a}{R}\right) +\frac{2\phi_1(\phi_1^{R/a}-1)}{\eta^2}\left(\frac{a}{R}\right)^2, & R\le a \\
2\phi_1\left(\frac{\phi_2}{\eta} - \phi_1 \right)\left(\frac{a}{R}\right) +
\phi_1\left(\phi_1 + \frac{2}{\eta}\left( \phi_1 - \frac{\phi_2}{\eta} \right) \right) \left(\frac{a}{R}\right)^2, & R\ge a.
\end{cases}
\end{align}
The large-$R$ asymptotic coefficients in distance units $a$ are given by
\begin{equation}
    \bar{A}_{_V} = 2\phi_1\left(\frac{\phi_2}{\eta} - \phi_1 \right), \qquad \bar{B}_{_V} = \phi_1\left(\phi_1 + \frac{2}{\eta}\left( \phi_1 - \frac{\phi_2}{\eta} \right) \right).
\end{equation}
The integral of the variance from $R=0$ to $R=L$ is given by
\begin{align}
        \Sigma_{_V}(L) = af(\phi_1) + ag(\phi_1)\frac{(a-L)}{L} + ah(\phi_1)\ln\left( \frac{a}{L} \right),
    \end{align}
where
    \begin{align}
        f(\phi_1) = -\phi_1^2 + \frac{2\phi_1\ln(\eta)}{\eta} - \frac{2\phi_1\mbox{Ei}(-\eta)}{\eta} + \frac{2(\gamma-1)\phi_1}{\eta} + \frac{2\phi_1}{\eta^2} - \frac{2\phi_1^2}{\eta^2},
    \end{align}
    \begin{align}
        g(\phi_1) = \frac{\phi_1\left[2\phi_2 - \phi_1(2+\eta)\eta\right]}{\eta^2},
    \end{align}
    \begin{align}
        h(\phi_1) = \frac{2\phi_1(\phi_1\eta-\phi_2)}{\eta}.
    \end{align}

For 1D equilibrium hard rods, the leading order large-$R$ asymptotic coefficient is given exactly by 
\begin{equation}
    \bar{A}_{_V} = \phi_2(1-\phi_2)^2,
\end{equation}
which is obtained from (\ref{A-packing}) and the fact that $S(0)=(1-\phi_2)^2$ \cite{To21c}.
The  large-$R$ asymptotic coefficient $\bar{B}_{_V}$ and integral $\Sigma_{_V}(L)$ are computed numerically. However, using the exact low-$\phi_2$ asymptotic expansion of $\bar{B}_{_V}$ (which is easily obtained) and the condition that $\bar{B}_{_V}$ must vanish at $\phi_2=1$, a fit of the numerical data using a polynomial of degree six (without up to quadratic terms) yields the highly accurate approximation formula in distance units $a$ for $\bar{B}_{_V}$ for all $\phi_2$, namely,
\begin{equation}
{\bar B}_{V}=\frac{-\phi_2}{6}\left(2 -7\phi_2+8\phi_2^2-3\phi_2^3\right).
\end{equation}

For 1D disordered hyperuniform media, the local variance is given by
\begin{equation}
    \sigma_{_V}^2(R) = \frac{\phi_1\phi_2a^2}{4R^2}\Big\{1-[\sin(2R/a) + \cos(2R/a)]  \exp(-2R/a) \Big\}
\end{equation}
The large-$R$ asymptotic coefficients in distance units $a$ are given by
\begin{equation}
    \bar{A}_{_V} = 0, \qquad \bar{B}_{_V} = \frac{\phi_1\phi_2}{4}.
\end{equation}
The integral of the variance over all $R$ is given by
\begin{equation}
    \Sigma_{_V}(\infty) =  \frac{a\, \pi \phi_1\phi_2}{4}. 
\end{equation}

We consider 1D periodic rods of length $2a=\phi_2b$ arranged on the sites of the integer lattice $\mathbb{Z}$, where $b$ is the lattice spacing. The local variance 
was determined analytically in Ref. \cite{Qu99}, but here we present slightly more compact
formulas. Specifically, we write the local variance for this periodic medium as follows:
\begin{equation}
\sigma_{_V}^2(R) = \frac{1}{4}\left( G - G^2 - \frac{F}{3} \right)\left(\frac{F}{R}\right)^2,
\label{1D-periodic}
\end{equation}
where
\begin{align}
F =
\begin{cases}
\min\{\{2R\},2a\}, & 2a \le 1- \{2R\} \\
\min\{1-\{2R\},1-2a\}, & 2a \ge 1 - \{2R\},
\end{cases}
\end{align}
\begin{align}
G =
\begin{cases}
\max\{\{2R\},2a\}, & 2a \le 1- \{2R\} \\
\max\{1-\{2R\},1-2a\}, & 2a \ge 1 - \{2R\},
\end{cases}
\end{align}
and $\{x\}$ denotes the fractional part of $x$.

From relation (\ref{1D-periodic}), we can ascertain that the large-$R$ asymptotic coefficients in distance units $a$ are given by
\begin{equation}
\bar{A}_{_V}=0, \qquad \bar{B}_{_V}= \frac{\phi_1^2}{6}.
\end{equation}
We also find that the integral of the variance over all $R$ is given by the following infinite sum:
\begin{equation}
\Sigma_{_V}(\infty) = \frac{a}{\pi^2\phi_2}\sum_{n=1}^{\infty}\frac{\sin^2(n\pi\phi_2)}{n^3}\le   \frac{a\,\zeta(3)}{\pi^2\phi_2},
\label{series}
\end{equation}
where $\zeta(x)$ is the Riemann zeta function. The upper bound on $\Sigma_{_V}(\infty)$ immediately follows by setting the numerator in (\ref{series})
to its maximal value of unity.  
The quantity  $\Sigma_{_V}(\infty)$ is a function of $\phi_2$ that is symmetric about $\phi_2=1/2$, where it is achieves its maximal value,
and equal to zero at the extreme values of $\phi_2$, i.e., $\phi_2=0$ and $\phi_2=1$.
It is noteworthy that when $\phi_2$ is a rational number such that $\sin^2(\pi\phi_2)$ is a rational
number, the infinite sum can be exactly represented entirely in terms of $\zeta(3)=1.202056903\ldots$. 
Specifically, when $\phi_2$ equals  $1/2$, $1/3$, $1/4$, and $1/6$,
$\Sigma_{_V}(\infty)$ equals $7\zeta(3)/(8\pi^2)$, $13\zeta(3)/(18\pi^2)$, $35\zeta(3)/(64\pi^2)$, $\zeta(3)/(3\pi^2)$, respectively. Whenever
$\sin^2(\pi\phi_2)$ is irrational, the infinite sum must be approximated by a finite sum,
but the resulting value is highly accurate because (\ref{series}) converges rapidly.

\begin{figure*}[!htp]
\centering
\includegraphics[width=0.6\textwidth,clip=]{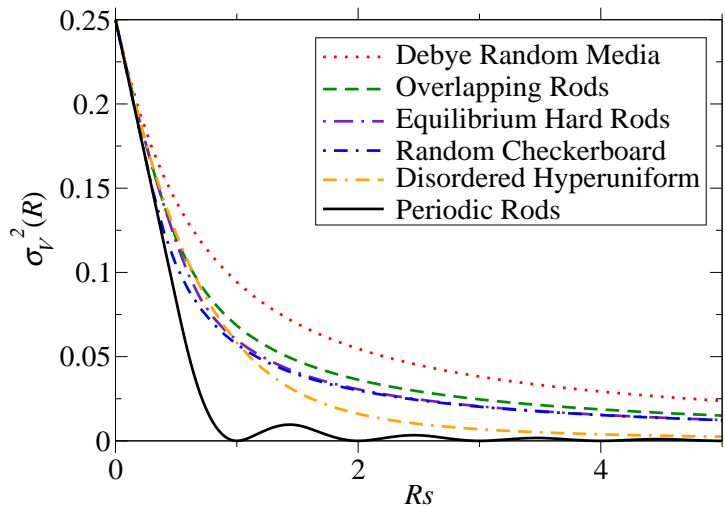}
\caption{Comparison of the local volume-fraction variance $\sigma_{_V}^2(Rs)$ versus the dimensionless
window radius $Rs$ for 1D models at $\phi_2=0.5$, where $s$ is the specific surface. }\label{1D-variances}
\end{figure*}

Table \ref{1D-Av-Bv} lists closed-form formulas and numeric values of the asymptotic coefficients $\bar{A}_{_V}$ and $\bar{B}_{_V}$ for the 1D models. The ``volume" coefficient ${\bar A}_{_V}$, which measures order at large length scales for nonhyperuniform media, is largest for Debye random media among all models considered and substantially larger for most values of $\phi_2$ than ${\bar A}_{_V}$ for overlapping rods, which is the second largest.
The coefficient ${\bar A}_{_V}$ for the random checkerboard and equilibrium rods are identical and the smallest
among the nonhyperuniform models. We note that when $\phi_2$ is sufficiently large,
${\bar B}_{_V}$ for equilibrium hard rods becomes positive (reflecting strong correlations), which indicates that this system becomes more ordered than the random checkerboard, flipping the ranking indicated in Table \ref{1D-Av-Bv}. For the hyperuniform models, we see that ${\bar B}_{_V}$ is smaller for periodic rods than it is for disordered hyperuniform media which is consistent with intuition. In summary, when ranking the degree of order of different nonhyperuniform media with asymptotic coefficients, it is best to use $\bar{A}_{_V}$ as this coefficient weights the dominant term in the large-$R$ asymptotic expansion of the variance for these systems. Analogously, one should use the coefficient $\bar{B}_{_V}$ when ranking order among hyperuniform media.

Figure \ref{1D-variances} compares plots of the local volume-fraction variance $\sigma^2_{_V}(Rs)$ versus the dimensionless window radius $Rs$ for 1D models where $\phi_2=0.5$. We see that local volume-fraction fluctuations for all window radii are bounded from above by Debye random media and from below by periodic rods. Table \ref{1D-table} provides values of the local volume-fraction variance $\sigma_{_V}^2(Rs)$ for selected values of the dimensionless window radius $Rs$. For almost all values of $Rs$, the ranking of the models is in the order indicated in Figure \ref{1D-variances} and Table \ref{1D-Av-Bv}, i.e., Debye random media is the most disordered, whereas periodic rods are the most ordered. It is noteworthy that the ranking ascertained from Figure \ref{1D-variances} and Table \ref{1D-table} is consistent with that predicted by the coefficients ${\bar A}_{_V}$ and ${\bar B}_{_V}$ for the nonhyperuniform and hyperuniform models, respectively.
Interestingly, for $Rs \gtrsim 1$, the local variances for equilibrium hard rods and the random checkerboard
are essentially identical. For  $Rs \lesssim 1$, the random checkerboard
is the second most ordered microstructure due to a greater degree of clustering on the underlying
integer lattice at those length scales, which reduces volume-fraction fluctuations relative to
those of hard rods. As expected, the disordered hyperuniform model is the second most ordered microstructure for almost all length scales ($Rs \gtrsim 1$).
Table \ref{1D-table} also lists the values of $\Sigma_{_V}(\infty)$ for the models. While
$\Sigma_{_V}(\infty)$  is divergent for all 1D nonhyperuniform media, for reasons noted
above, the magnitude of $\Sigma_{_V}(\infty)$ for the hyperuniform models
provides an integral measure over all length scales that is consistent
with almost all local values of the variance, except when $Rs$ is very small.

\begin{table*}
\caption{Comparison of the asymptotic coefficients $\bar{A}_{_V}$ and $\bar{B}_{_V}$ for the 1D models. The microscopic characteristic length scale $D$ is chosen to be $s^{-1}$, where $s$ is the specific surface. }
\label{1D-Av-Bv}
\hspace{-1cm}\begin{tabular}{| c | c c c c |}
\hline
Model & $\bar{A}_{_V}$ & $\bar{B}_{_V}$ & $\bar{A}_{_V}(\phi_1=\frac{1}{2})$ & $\bar{B}_{_V}(\phi_1=\frac{1}{2})$ \\ [0.5ex]
\hline
Debye Random Media & $2(\phi_1\phi_2)^2$ & $-2(\phi_1\phi_2)^3$ & $0.12500$ & $-0.03125$\\
\hline
Overlapping Rods &  $2\phi_1^2(\phi_2-\phi_1\eta)$ & $\phi_1^3(\phi_1\eta^2 + 2\phi_1\eta - 2\phi_2)$  & $0.07671$ & $-0.00833$\\
\hline
Equilibrium Hard Rods & $(\phi_1\phi_2)^2$  & 
$-\frac{\phi_2^3}{6}\left(2 -7\phi_2+8\phi_2^2-3\phi_2^3\right)$  & $0.06250$ & $-0.00260$ \\
\hline
Random Checkerboard & $(\phi_1\phi_2)^2$ & $-\frac{1}{3}(\phi_1\phi_2)^3$  & $0.06250$ & $-0.00521$ \\
\hline
Disordered Hyperuniform & $0$ & $4(\phi_1\phi_2)^3$  & $0$ & $0.06250$\\
\hline
Periodic Rods & $0$ & $\frac{1}{6}(\phi_1\phi_2)^2$  & $0$ & $0.01042$\\ [1ex]
\hline
\end{tabular}
\end{table*}

\begin{table*}
\caption{
Comparison of the local volume-fraction variance $\sigma_{_V}^2(Rs)$ for 1D
models at $\phi_2=0.5$ for selected values of the dimensionless window
radius $Rs$, where $s$ is the specific surface. Included in the table is
the value
of the integral $\Sigma_{_V}(\infty)$.}
\label{1D-table}
\hspace{-2cm}\begin{tabular}{|c | c c c c c c | c |}\hline
1D Model & $\sigma_{_V}^2(0.3)$ & $\sigma_{_V}^2(1)$ &
$\sigma_{_V}^2(4.5)$ & $\sigma_{_V}^2(9.5)$ & $\sigma_{_V}^2(50)$ &
$\sigma_{_V}^2(100)$ & $\Sigma_{_V}(\infty)$\\ [0.5ex]
\hline
Debye Random Media & $0.17403$ & $0.09432$ & $0.02623$ & $0.01281$ &
$0.00249$ & $0.00125$ & $+\infty$ \\
\hline
Overlapping Rods & $0.16337$ & $0.06838$ & $0.01664$ & $0.00798$ &
$0.00153$ & $0.00077$ & $+\infty$ \\
\hline
Equilibrium Hard Rods & $0.16337$ & $0.05974$ & $0.01376$ & $0.00655$ &
$0.00125$ & $0.00062$ & $+\infty$ \\
\hline
Random Checkerboard & $0.15046$ & $0.05729$ & $0.01363$ & $0.00652$ &
$0.00125$ & $0.00062$ & $+\infty$ \\
\hline
Disordered Hyperuniform & $0.16470$ & $0.05833$ & $0.00309$ & $0.00069$ &
$0.00002$ & $0.00001$ & $0.19635$ \\
\hline
Periodic Rods & $0.15000$ & $0.00000$ & $0.00103$ & $0.00023$ & $0.00000$
& $0.00000$ & $0.10645$ \\ [1ex]
\hline
\end{tabular}
\end{table*}

\begin{figure*}[!htp]
\centering
\includegraphics[width=0.6\textwidth,clip=]{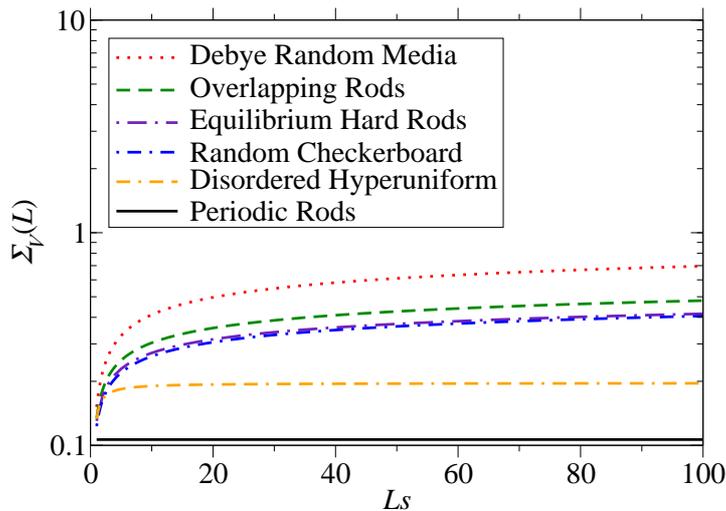}
\caption{Comparison of the integral of the local volume-fraction variance $\Sigma_{_V}(Ls)$ versus the dimensionless
distance $Ls$ for 1D models at $\phi_2=0.5$, where $s$ is the specific surface. }\label{1D-integrals}
\end{figure*}

Figure \ref{1D-integrals} compares plots of the integral of the local volume-fraction variance $\Sigma_{_V}(Ls)$ versus the dimensionless
window radius $Rs$ for 1D models where $\phi_2=0.5$. The rate of growth of $\Sigma_{_V}(Ls)$ with the dimensionless length $Ls$
for all nonhyperuniform media is consistent with the aforementioned ranking of the 1D models
for the local variance indicated in Figure \ref{1D-variances}, Table \ref{1D-Av-Bv}, and Table \ref{1D-table}.
Of course, $\Sigma_{_V}(Ls)$ achieves a constant asymptotic value for the hyperuniform models in the limit $Ls\to \infty$. 

\subsection{2D Models}

We present exact closed-form formulas for $\sigma^2_{_V}(Rs)$ and   $\Sigma_{_V}(\infty)$ for Debye random media.
For the remaining seven 2D models described in Sec. \ref{models}, we compute the same quantities 
for a volume fraction of phase 2 $\phi_2=0.4$, which is chosen because this is nearly the highest
value of $\phi_2$ consistent with a disordered stealthy hyperuniform packing. Here we also consider stealthy hyperuniform packings, which
we did not examine in one dimension. In the case of 2D periodic packings, we study the effect
of particle shape on the degree of order by considering both circular disks and hexagons
on the sites of the triangular lattice.

For a 2D Debye random medium, the local variance is given by
\begin{equation}
    \sigma_{_V}^2(R) =  \phi_1\phi_2 \Bigg\{\left[ 4I_0(2R/a) - 4L_0(2R/a)+2 \right]\left(\frac{a}{R} \right)^2 + 6\left[-I_1(2R/a)+L_1(2R/a) \right]\left(\frac{a}{R}\right)^3\Bigg\},
\end{equation}
where $I_{\alpha}(x)$ is the modified Bessel function of the first kind of order $\alpha$, and $L_{\alpha}(x)$ is the modified Struve function of order $\alpha$. The large-$R$ asymptotic coefficients in distance units of $a$ are given by
\begin{equation}
 \bar{A}_{_V} = 2\phi_1\phi_2, \qquad \bar{B}_{_V} = - \frac{8\phi_1\phi_2}{\pi}.
\end{equation}
The integral of the variance over all $R$ is given by
\begin{equation}
\Sigma_{_V}(\infty)=\frac{16}{3\pi}.
\end{equation}

\begin{figure*}[!htp]
\centering
\includegraphics[width=0.6\textwidth,clip=]{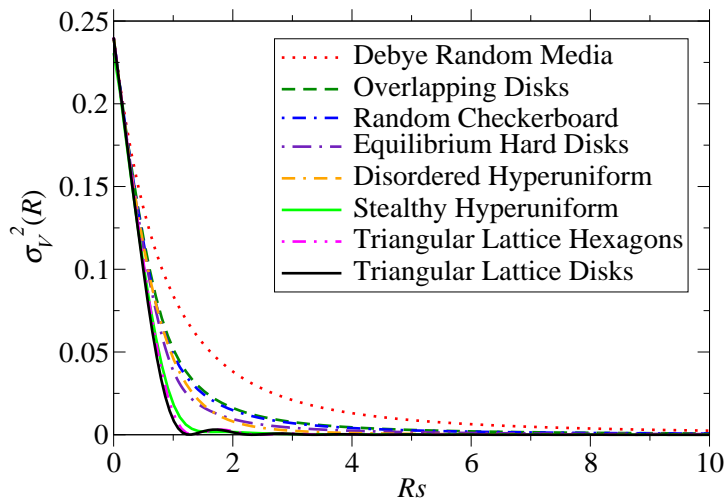}
\caption{Comparison of the local volume-fraction variance $\sigma_{_V}^2(Rs)$ versus the dimensionless
window radius $Rs$ for 2D models at $\phi_2=0.4$, where $s$ is the specific surface. }\label{2D-variances}
\end{figure*}

Table \ref{2D-Av-Bv} lists numeric values of the asymptotic coefficients $\bar{A}_{_V}$ and $\bar{B}_{_V}$ for the 2D models at $\phi_2=0.40$. Here, the nonhyperuniform and hyperuniform media are listed in order of decreasing $\bar{A}_{_V}$ and $\bar{B}_{_V}$, respectively, reflecting the ranking of the degree of order in these systems by the asymptotic coefficients. Figure \ref{2D-variances} compares plots of the local volume-fraction variance $\sigma^2_{_V}(Rs)$ versus the dimensionless window radius $Rs$ for the 2D models at $\phi_2=0.4$. Similar to what was observed in 1D, we see that the local volume-fraction variance for the 2D models are bounded from above by Debye random media for all window radii and bounded from below by the triangular lattice of circular disks for most window radii. Table \ref{2D-table} provides values of the local volume-fraction variance $\sigma_{_V}^2(Rs)$ for selected values of the dimensionless window radius $Rs$. For almost all values of $Rs$, the ranking of the models is in the order indicated in Fig. \ref{2D-variances} and Tables \ref{2D-Av-Bv} and \ref{2D-table}, i.e., Debye random media is the most disordered, down to the triangular lattice of circular disks, which is the most ordered for reasons noted in Sec. \ref{specs}. While overlapping disks is the second most disordered model (as it is for the 1D models considered), the random checkerboard is more disordered than equilibrium hard disks for almost all length scales ($Rs \gtrsim 1$), reversing the general trend observed for their one-dimensional counterparts (see Fig. \ref{1D-variances} and Table \ref{1D-table}). This reversal of rank ordering highlights the effect of dimensionality on the degree of order for the same models and occurs for reasons given in Sec. \ref{specs}, i.e., the spectral density for equilibrium hard disks is lower than that of the random checkerboard for small wavenumbers (see Fig. \ref{2D-spectral}). The fact that the stealthy packing is the most ordered among the disordered models was also explained in Sec. \ref{specs}. Lastly, we note that the ranking of order via the integrated variance $\Sigma_{_V}(\infty)$ in the rightmost column of Table \ref{2D-table} agrees with that given by both the asymptotic coefficients as well as the values of the local variance $\sigma_{_V}^2(Rs)$ for very large (e.g., $Rs>100$) length scales.

\begin{table*}
\caption{Comparison of the asymptotic coefficients $\bar{A}_{_V}$ and $\bar{B}_{_V}$ 
for the 2D models at $\phi_2=0.4$. The microscopic characteristic length scale $D$ is chosen to be $s^{-1}$, where $s$ is the specific surface.}
\label{2D-Av-Bv}
\begin{tabular}{| c | c c |}
\hline
Model & $\bar{A}_{_V}$ & $\bar{B}_{_V}$\\ [0.5ex]
\hline
Debye Random Media & $0.27287$ & $-0.26196$ \\
\hline
Overlapping Disks & $0.07808$ & $-0.02679$ \\
\hline
Random Checkerboard & $0.07041$ & $-0.02244$ \\
\hline
Equilibrium Hard Disks & $0.03827$ & $-0.00208$ \\
\hline
Disordered Hyperuniform & $0$ & $0.06549$ \\
\hline
Stealthy Hyperuniform & $0$ & $0.01183$ \\
\hline
Triangular Lattice Hexagons & $0$ & $0.00965$ \\
\hline
Triangular Lattice Disks & $0$ & $0.00844$ \\
\hline
\end{tabular}
\end{table*}

\begin{table*}
\caption{
Comparison of the local volume-fraction variance $\sigma_{_V}^2(Rs)$ for 2D
models at $\phi_2=0.4$ for selected values of the dimensionless window
radius $Rs$, where $s$ is the specific surface. Included in the table is
the value
of the integral $\Sigma_{_V}(\infty)$.}
\label{2D-table}
\hspace{-2.5cm}\begin{tabular}{|c | c c c c c c | c |}
\hline
2D Model & $\sigma_{_V}^2(0.3)$ & $\sigma_{_V}^2(1)$ &
$\sigma_{_V}^2(4.5)$ & $\sigma_{_V}^2(9.5)$ & $\sigma_{_V}^2(50)$ & $\sigma_{_V}^2(100)$ & $\Sigma_{_V}(\infty)$\\ [0.5ex]
\hline
Debye Random Media & $0.16978$ & $0.08406$ & $0.01064$ & $0.00272$ & $0.00011$ & $0.00003$ & $0.30720$ \\
\hline
Overlapping Disks & $0.16164$ & $0.05186$ & $0.00356$ & $0.00083$ & $0.00003$ & $0.00001$ & $0.18730$ \\
\hline
Random Checkerboard & $0.16100$ & $0.04838$ & $0.00323$ & $0.00075$ & $0.00003$ & $0.00001$ & $0.18509$ \\
\hline
Equilibrium Hard Disks & $0.23145$ & $0.03878$ & $0.00166$ & $0.00042$ & $0.00002$ & $3.733\times10^{-6}$ & $0.15880$ \\
\hline
Disordered Hyperuniform & $0.15831$ & $0.04616$ & $0.00071$ & $0.00008$ & $5.239\times10^{-7}$ & $6.549\times10^{-8}$ & $0.15360$ \\
\hline
Stealthy Hyperuniform& $0.15528$& $0.00194$ & $0.00012$& $0.00001$& $9.372\times10^{-8}$& $9.541\times10^{-9}$& $0.11890$ \\
\hline
Triangular Lattice Hexagons & $0.15661$ & $0.01319$ & $0.00003$ & $0.00002$ & $1.351\times10^{-7}$ & $6.721\times10^{-10}$ & $0.11523$ \\
\hline
Triangular Lattice Disks & $0.15426$ & $0.00915$ & $0.00001$ & $3.758\times10^{-6}$ & $9.588\times10^{-8}$ & $1.480\times10^{-8}$ & $0.11071$ \\ [1ex]
\hline
\end{tabular}
\end{table*}

\subsection{3D Models}

We present exact closed-form formulas for $\sigma^2_{_V}(R)$ and  $\Sigma_{_V}(L)$ or  $\Sigma_{_V}(\infty)$
for three of the eight 3D models described in Sec. \ref{models}: antihyperuniform media, Debye random media, and 
disordered hyperuniform media. In all other cases, we compute the same quantities 
for a phase 2 volume fraction  $\phi_2=0.38$, which is chosen because this is nearly the highest
value of $\phi_2$ consistent with a disordered stealthy packing. We also consider antihyperuniform media, which was not done in the lower
dimensions. In the case of 3D periodic packings of spheres, we study the effect
of the lattice on the degree of order by considering both the SC and BCC arrangements.

For the 3D antihyperuniform model defined in Sec. \ref{models}, the local variance is given by
\begin{eqnarray}
    \sigma_{_V}^2(R) &=&     \frac{9\phi_1\phi_2}{4}\left(\frac{a}{R} \right)^2 +  \frac{\phi_1\phi_2}{16}\left[176 - 96\ln\left(\frac{2R}{a}+1\right) \right]\left(\frac{a}{R} \right)^3 \nonumber\\ 
&+& \frac{\phi_1\phi_2}{16} \left[30-108\ln\left(\frac{2R}{a}+1\right)\right]\left(\frac{a}{R} \right)^4 - \frac{15\phi_1\phi_2}{8}\left(\frac{a}{R} \right)^5 \nonumber \\
&+& \frac{15\phi_1\phi_2}{16}\ln\left(\frac{2R}{a}+1\right)\left(\frac{a}{R} \right)^6 
\end{eqnarray}
Notice that for large $R$, the variance has the scaling $\sigma_{_V}^2(R)\sim R^{-2}$, which is clearly slower than window-volume decay (i.e., $R^{-3}$), 
obeyed by typical nonhyperuniform media. Thus, the asymptotic coefficients $\bar{A}_{_V}$ and $\bar{B}_{_V}$ do not exist, i.e., they are unbounded.
The integral of the variance from $R=0$ to $R=L$ is given by
\begin{eqnarray}
    \Sigma_{_V}(L)   &=& \frac{6a\phi_1\phi_2}{5}-\frac{9a\phi_1\phi_2}{4}\left(\frac{a}{L}\right)+ a\phi_1\phi_2\Big[3\ln\left(\frac{2L}{a}+1\right)-4 \Big]\left(\frac{a}{L}\right)^2 \nonumber \\
&+& \frac{9a\phi_1\phi_2}{4}\left[\ln\left(\frac{2L}{a}+1\right)-\frac{1}{6} \right]\left(\frac{a}{L} \right)^3 
\frac{3a\phi_1\phi_2}{8}\left(\frac{a}{L}\right)^4   - \frac{3a\phi_1\phi_2}{16}\ln\left(\frac{2L}{a}+1 \right) \left(\frac{a}{L}\right)^5 .\nonumber\\
\end{eqnarray}

For 3D Debye random media, the local variance is given by
\begin{align}
    \sigma_{_V}^2(R) =  \phi_1\phi_2 \Bigg\{3\left[2 - 3\exp(-2R/a) \right]\left(\frac{a}{R} \right)^3 - \frac{9}{2}\left[3 + 7\exp(-2R/a) \right]\left(\frac{a}{R} \right)^4 \nonumber \\
    - 45\exp(-2R/a)\left( \frac{a}{R} \right)^5 + \frac{45}{2}\left[1-\exp(-2R/a)\right]\left( \frac{a}{R} \right)^6\Bigg\}.
\end{align}
The large-$R$ asymptotic coefficients in distance units of $a$ are given by
\begin{equation}
 \bar{A}_{_V} = 6\phi_1\phi_2, \qquad \bar{B}_{_V} = - \frac{27\phi_1\phi_2}{2}.
\end{equation}

The integral of the variance over all $R$ is given by
\begin{equation}
\Sigma_{_V}(\infty)=\frac{6\phi_1\phi_2 a}{5}.
\label{Sigma-3D-Debye}
\end{equation}

The local variance of 3D disordered hyperuniform media is given by
\begin{eqnarray}
    \sigma_{_V}^2(R) &=& \phi_1\phi_2\exp\left( \frac{-2R}{a} \right)\Bigg\{ \left[\frac{27\sqrt{3}}{8}\sin\left( \frac{2R}{a\sqrt{3}} \right) \right]\left( \frac{a}{R} \right)^3 \nonumber \\
    &+&\frac{1}{64}\left[243\exp\left(\frac{2R}{a}\right) + 567\sqrt{3}\sin\left( \frac{2R}{a\sqrt{3}} \right) + 567\cos\left( \frac{2R}{a\sqrt{3}} \right)\right]\left( \frac{a}{R} \right)^4 \nonumber \\
    &+&\frac{1}{64}\left[ 405\sqrt{3}\sin\left( \frac{2R}{a\sqrt{3}} \right) + 1215\cos\left( \frac{2R}{a\sqrt{3}} \right) \right]\left( \frac{a}{R} \right)^5 \nonumber \\
    &+& \frac{1215}{128}\left[ \cos\left( \frac{2R}{a\sqrt{3}} \right) - \exp\left(\frac{2R}{a}\right) \right]\left( \frac{a}{R} \right)^6 \Bigg\}
\end{eqnarray}
The large-$R$ asymptotic coefficients in distance units $a$ are given by
\begin{equation}
    \bar{A}_{_V} = 0, \qquad \bar{B}_{_V} = \frac{243\phi_1\phi_2}{64}.
\end{equation}
The integral of the variance over all $R$ is given by
\begin{equation}
    \Sigma_{_V}(\infty) = \frac{9\phi_1\phi_2a}{10},
\end{equation}
which, as expected, is smaller than that for 3D Debye random media [cf. (\ref{Sigma-3D-Debye})].

\begin{figure*}[!htp]
\centering
\includegraphics[width=0.6\textwidth,clip=]{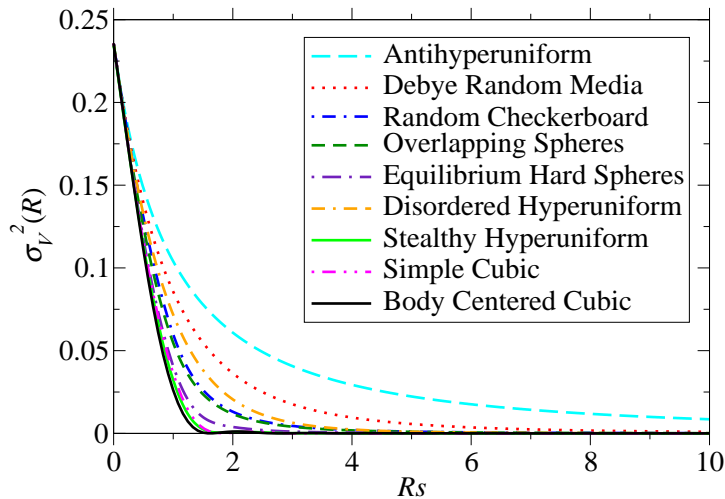}
\caption{Comparison of the local volume-fraction variance $\sigma_{_V}^2(Rs)$ versus the dimensionless
window radius $Rs$ for 3D models at $\phi_2=0.38$, where $s$ is the specific surface.}\label{3D-variances}
\end{figure*}

Table \ref{3D-Av-Bv} provides values of the asymptotic coefficients $\bar{A}_{_V}$ and $\bar{B}_{_V}$ for the 3D models at $\phi_2=0.38$. Once again, the nonhyperuniform and hyperuniform media are listed in order of decreasing $\bar{A}_{_V}$ and $\bar{B}_{_V}$, respectively. Notably, $\bar{B}_{_V}$ ranks the disordered stealthy hyperuniform packing as more ordered than the crystalline simple cubic one. Figure \ref{3D-variances} compares plots of the local volume-fraction variance $\sigma^2_{_V}(Rs)$ versus the dimensionless window radius $Rs$ for the 3D models at $\phi_2=0.38$. From this plot, we see that local volume-fraction variances in the 3D models are bounded from above by antihyperuniform media for all window radii and bounded from below by body-centered cubic lattice of spheres for nearly all window radii. Table \ref{3D-table} provides values of the local volume-fraction variance $\sigma_{_V}^2(Rs)$ for selected values of the dimensionless window radius $Rs$. For almost all values of $Rs$, the ranking of the models is in the order indicated in Figure \ref{3D-variances} and Table \ref{3D-table}, i.e., antihyperuniform media is the most disordered, followed by Debye random media, down to the BCC lattice of spheres, which is the most ordered for reasons noted in Sec. \ref{specs}. However, ranking order according to the integrated variance $\Sigma_{_V}(\infty)$ predicts that disordered hyperuniform media lies between Debye random media and the random checkerboard. Unlike what was observed in the 1D and 2D systems, this ranking of order contrasts that predicted by both the local variance at large length scales and the asymptotic coefficients $\bar{A}_{_V}$ and $\bar{B}_{_V}$ and can be explained by comparing the relative sizes of the spectral densities for these systems presented in Figure \ref{3D-spectral}.

As was observed in 1D and 2D, Debye random media is still the most disordered among all of the typical disordered nonhyperuniform [see Eq. (\ref{sigma-nonhyper})]  models considered here. We also note that, as was the case in 2D, the stealthy packing is the most ordered among the disordered models and the simple cubic lattice packing according to all three order metrics. In both cases, these results occur for reasons provided in Sec. \ref{specs}. Interestingly, the 3D random checkerboard is more disordered than overlapping spheres for all length scales, reversing the ranking of their two-dimensional counterparts (see Figure \ref{2D-variances} and Table \ref{2D-table}). Overall, we see that the random checkerboard becomes progressively more disordered relative to the other models as the spatial dimension increases, again, highlighting the effect of dimensionality on the degree of order for a given system. This growing disorder with $d$ in the random checkerboard model can be attributed to a generalized decorrelation principle \cite{To06b,To06d} (see Sec. \ref{conclusions} for more information).


\begin{table*}
\caption{Comparison of the asymptotic coefficients $\bar{A}_{_V}$ and $\bar{B}_{_V}$ 
for the 3D models at $\phi_2=0.38$. The microscopic characteristic length scale $D$ is chosen to be $s^{-1}$, where $s$ is the specific surface.}
\label{3D-Av-Bv}
\begin{tabular}{| c | c c |}
\hline
Model & $\bar{A}_{_V}$ & $\bar{B}_{_V}$\\ [0.5ex]
\hline
Antihyperuniform Media & - & - \\
\hline
Debye Random Media & $1.18313$ & $-2.50871$ \\
\hline
Random Checkerboard & $0.15888$ & $-0.11146$ \\
\hline
Overlapping Spheres & $0.13994$ & $-0.09422$ \\
\hline
Equilibrium Hard Spheres & $0.02680$ & $0.01765$ \\
\hline
Disordered Hyperuniform & $0$ & $0.70557$ \\
\hline
Simple Cubic & $0$ & $0.01693$ \\
\hline
Stealthy Hyperuniform & $0$ & $0.01539$ \\
\hline
Body Centered Cubic & $0$ & $0.01171$ \\
\hline
\end{tabular}
\end{table*}

\begin{table*}
\caption{Comparison of the local volume-fraction variance $\sigma_{_V}^2(Rs)$ for 3D models at $\phi_2=0.38$ for selected values of the dimensionless window radius $Rs$,
where $s$ is the specific surface. Included in the table is the value of the integral $\Sigma_{_V}(\infty)$.}
\label{3D-table}
\hspace{-2.5cm}\begin{tabular}{|c | c c c c c c | c |}
\hline
Model & $\sigma_{_V}^2(0.3)$ & $\sigma_{_V}^2(1)$ & $\sigma_{_V}^2(4.5)$ & $\sigma_{_V}^2(9.5)$ & $\sigma_{_V}^2(50)$ & $\sigma_{_V}^2(100)$ & $\Sigma_{_V}(\infty)$\\
[0.5ex]
\hline
Antihyper. Media & $0.17538$ & $0.10396$ & $0.02545$ & $0.00921$ & $0.00057$ & $0.00016$ & $\infty$ \\
\hline
Debye Random Media & $0.17104$ & $0.08567$ & $0.00731$ & $0.00108$ & $9.064\times10^{-6}$ & $1.158\times10^{-6}$ & $0.26644$ \\
\hline
Random Checkerboard & $0.16813$ & $0.05982$ & $0.00147$ & $0.00017$ & $1.246\times10^{-6}$ & $1.568\times10^{-7}$ & $0.18015$ \\
\hline
Overlapping Spheres & $0.16456$ & $0.05471$ & $0.00131$ & $0.00015$ & $1.105\times10^{-6}$ & $1.391\times10^{-7}$ & $0.17132$ \\
\hline
Equil. Hard Spheres & $0.16098$ & $0.04070$ & $0.00032$ & $0.00003$ & $2.163\times10^{-7}$ & $2.695\times10^{-8}$ & $0.14226$ \\
\hline
Disordered Hyper. & $0.16785$ & $0.07169$ & $0.00153$ & $0.00008$ & $1.128\times10^{-7}$ & $7.054\times10^{-9}$ & $0.19983$ \\
\hline
Stealthy Hyper. & $0.15874$ & $0.03197$ & $0.00003$ &	$1.869\times 10^{-6}$&	$2.492\times10^{-9}$ & $1.450\times10^{-10}$& $0.12795$ \\ 
\hline
Simple Cubic & $0.15961$ & $0.03631$ & $7.524\times10^{-6}$ & $6.757\times10^{-8}$ & $4.653\times10^{-10}$ & $2.467\times10^{-10}$ & $0.13176$ \\
\hline
Body Centered Cubic & $0.15916$ & $0.02602$ & $0.00005$ & $2.320\times10^{-7}$ & $1.920\times10^{-9}$ & $4.859\times10^{-12}$ & $0.12327$ \\ [1ex]
\hline
\end{tabular}
\end{table*}

\section{Conclusions and Discussion}
\label{conclusions}

In this work, we have taken initial steps to devise order metrics to characterize the microstructures of disordered and ordered two-phase media 
across all length scales via the local volume-fraction variance $\sigma^2_{_V}(R)$. By studying
a total of 22 two-phase models
across the first three space dimensions, including those that span from nonhyperuniform
and hyperuniform ones with varying degrees of short- and long-range order,
we found that $\sigma^2_{_V}(R)$ as a function of the dimensionless window radius $Rs$ provides
a reasonably robust and sensitive order metric across length scales. Additionally, we determined that the asymptotic coefficients $\bar{A}_{_V}$ and $\bar{B}_{_V}$ as well as the integrated volume-fraction variance $\Sigma_{_V}(\infty)$ are similarly effective order metrics. To compare the degree of disorder for different microstructures at a fixed volume fraction and at a specific length scale $\ell$, the local volume-fraction variance $\sigma_V^2(\ell)$ should be used. To make such comparisons at larger length scales, the asymptotic coefficients $\bar{A}_{_V}$ or $\bar{B}_{_V}$, for nonhyperuniform or hyperuniform media, respectively, is a reasonable and natural choice. Lastly, for an overall quantification of disorder across all length scales in a system, the integrated variance could be used.

Interestingly, using all three metrics, Debye random media is the most disordered among all of the typical disordered nonhyperuniform [see Eq. (\ref{sigma-nonhyper})] models examined in this work across all three dimensions.
In two and three dimensions, we found that the stealthy disordered hyperuniform sphere packing is the most ordered among all disordered models considered. An important lesson learned from our study is that the relative order of any particular $d$-dimensional model can change with $d$. For example, going from 1D to 3D, the disordered hyperuniform medium becomes progressively more disordered at short length scales (even if more ordered at intermediate and large length scales)--having a volume fraction variance comparable to that of the random checkerboard in 3D for smaller window radii. The random checkerboard also becomes progressively more disordered relative to the other systems as the dimension increases. Note that the number of neighbors for a cell in the $d$-dimensional random checkerboard is given by $3^d-1$. Therefore, as $d$ increases, the number of potential directions for pair correlations increases exponentially, reducing the overall likelihood of spatial correlations in this model. This specific higher-dimensional behavior is a manifestation of the decorrelation principle \cite{To06b,To06d}.

It is important to recognize that the ranking of order/disorder of two-phase microstructures via
the local variance at fixed phase volume fraction in any particular dimension depends
on the choice of the microscopic characteristic length scale $D$. We have chosen  $D$ to be 
the inverse of the specific surface because it is broadly applicable
and easily determined \cite{Ki21}. An interesting topic for future research
is the search and evaluation of a length scale $D$ that is superior to $s^{-1}$
for improving the rank order of two-phase microstructures. An obvious extension of the present work
is the formulation of order metrics to $n$-phase
media, which is formally straightforward.

To devise a metric that follows previous considerations for point configurations \cite{To18b} so that a scalar order metric $\psi$ for two-phase media takes on the value $1$ for the most ordered state and $0$ for the most disordered state for length scales between $L_1$ and $L_2$, a reasonable choice that could be used is the following ratio:
\begin{equation}
    \psi = \frac{\displaystyle \min_{\mathcal{C}}\{\Sigma_{_V}(L_2)-\Sigma_{_V}(L_1) \}}{ \Sigma_{_V}(L_2) -\Sigma_{_V}(L_1)},
\end{equation}
where $\Sigma_{_V}(L)$ is defined in (\ref{Sigma-L}) and $\mathcal{C}$ denotes the set of all two-phase media. Since the determination of the minimum is a notoriously difficult task for which there are no proofs, in practice, the minimum is determined from all candidate structures. Finally, we note a relationship between the recently introduced concept of ``spreadability" for time-dependent diffusion in two-phase media \cite{To21d,Ma22,Wa22} and the order metric as measured by $\vv{R}$. The spreadability was determined for a subset
of the models studied here across dimensions. For this common subset of models,
the spreadability at short, intermediate, and long times is roughly proportional to the magnitude of the local variance at short, intermediate, and long length scales, respectively, 
thus registering the same rankings between these common models. Establishing the precise reasons for this link between the spreadability and the local variance more rigorously
is an outstanding problem for future research.

\bigskip
The authors gratefully acknowledge the
support of the Air Force Office of Scientific Research Program on
Mechanics of Multifunctional Materials and Microsystems under
award No. FA9550-18-1-0514.

\appendix

\section{General Asymptotic Analysis}
\label{asy-gen}

The local volume-fraction variance $\sigma^2_{_V}(R)$ is generally a function
that can be decomposed into a global part that decreases with
the window radius $R$ and a local part that oscillates on microscopic
length scales about the global contribution. The more general large-$R$ asymptotic formula for the variance for a large class of statistically homogeneous media is given by \cite{To18a}:
\begin{equation}\label{eq:large-R_exp}
\sigma_{_V}^2(R) = A_{V}(R) \qty(\frac{D}{R})^d + B_{V}(R) \qty(\frac{D}{R})^{d+1} + o\qty(\frac{D}{R})^{d+1},
\end{equation}
where 
\begin{align}
A_{V}(R) =& \frac{1}{\fn{v_1}{D}} \int_{\abs{\vect{r}}\leq 2R} \fn{\chi_{_V}}{\vect{r}}\dd{\vect{r}} \\
B_{V}(R)=& -\frac{\fn{c}{d}}{2D\fn{v_1}{D}} \int_{\abs{\vect{r}}\leq 2R} \fn{\chi_{_V}}{\vect{r}}\abs{\vect{r}}\dd{\vect{r}},
\end{align}
are $R$-dependent coefficients.   Observe that when the  coefficients $A_{V}(R)$ and   $B_{V}(R)$
 converge in the limit $R \rightarrow \infty$, they are  equal
to the constants ${\bar A}_{V}(R)$ and   ${\bar B}_{V}(R)$, defined by (\ref{A}) and (\ref{B}), respectively. The more general asymptotic analysis has been applied to yield the scaling laws for class II and class III hyperuniform media indicated in relation \ref{sigma-hyper} \cite{To18a,To20}.
\bigskip

In cases when two-phase media are generated via simulations,
it is advantageous to estimate the coefficients $\bar{A}_{_V}$ and $\bar{B}_{_V}$ by using the cumulative moving average, as defined in Ref. \cite{To18a}, namely,
\begin{equation}\label{eq:Av-asy}
{\bar A}_{_V} \equiv \lim_{L\to\infty} \frac{1}{L} \int_0^L {\bar A}_{_V}(R) \dd{R},
\end{equation}
\begin{equation}\label{eq:Bv-asy}
{\bar B}_{_V} \equiv \lim_{L\to\infty} \frac{1}{L} \int_0^L {\bar B}_{_V}(R) \dd{R}.
\end{equation}

\providecommand{\newblock}{}

\end{document}